\documentclass[twocolumn,english,pra]{revtex4-1}
\usepackage[T1]{fontenc}
\usepackage[latin9]{inputenc}
\usepackage{color}
\usepackage{amsmath}
\usepackage{amssymb}
\usepackage{graphicx}

\makeatletter
\usepackage[colorlinks = true,
            linkcolor =red,
            urlcolor  =magenta,
            citecolor = blue,
            anchorcolor = blue]{hyperref}

\makeatother

\usepackage{babel}
\begin{document}
\title{Post-selected von Neumann Measurement with  Superpositions of Orbital-Angular-Momentum
Pointer States}
\author{Janarbek Yuanbek$^{1,2}$}
\author{Yi-Fang Ren$^{1}$}
\author{Yusuf Turek$^{1}$}
\email{yusuftu1984@hotmail.com}

\affiliation{$^{1}$School of Physics,Liaoning University,Shenyang,Liaoning 110036,China}
\affiliation{$^{2}$School of Physics and Electronic Engineering, Xinjiang Normal
University, Urumqi, Xinjiang 830054, China}
\date{\today}
\begin{abstract}
We investigated an orbital angular momentum (OAM) pointer within the
framework of von Neumann measurements and discovered its significant
impact on optimizing superpositions of Gaussian and Laguerre-Gaussian
(LG) states. Calculations of the quadrature squeezing, the second-order
cross-correlation function, the Wigner function, and the signal-to-noise
ratio (SNR) support our findings. Specifically, by carefully selecting
the anomalous weak value and the coupling strength between the measured
system and the pointer, we demonstrated that the initial Gaussian
state transforms into a non-Gaussian state after postselection. This
transition highlights the potential of OAM pointers in enhancing the
performance of quantum systems by tailoring state properties for specific
applications.
\end{abstract}
\maketitle

\section{Introduction}

A Laguerre-Gaussian (LG) beam combines Laguerre polynomials with a
Gaussian function and describes quantum states exhibiting spherical
symmetry. The LG mode, characterized by a zero-intensity central point
(ZIP), is called an optical vortex beam \citet{21-nye1974dislocations}.
Researchers have proposed various generation methods and experimental
implementations for these beams \citep{22-bazhenov1990laser,30-Kobayashi:12,23-doi:10.1080/09500349114552651,24-Heckenberg:92,25-PhysRevA.45.8185,26-BEIJERSBERGEN1994321,27-PhysRevLett.96.163905,28-Ando:09,29-PhysRevA.84.033813}.
In recent years, optical vortex beams have gained significant attention
due to their distinctive spatial structure, which is highly useful
in quantum information science \citep{Kim2013SelectiveGO,Xia2022UltrasensitiveMO}
for transmitting quantum states and generating quantum entanglement
\citep{qmen-RevModPhys.81.865,qmen-Vedral2014QuantumE,qmen-Tavares2023QuantumE}.
Allen \textit{et al.} \citep{AllenPhysRevA.45.8185} demonstrated
that the LG beam---expressed by a phase cross-section of $\exp(il\varphi)$,
where $l$ takes integer values---carries an orbital angular momentum
(OAM) of $l\hbar$. Here, $\varphi$ is the azimuthal angle, and $l$
is the topological charge (TC). OAM beams also have numerous potential
applications, including optical manipulation \citep{6-grier2003revolution},
optical communication \citep{8-Gibson:04,9-wang2012terabit,10-doi:10.1126/science.1237861,11-doi:10.1073/pnas.1612023113,12-doi:10.1126/sciadv.1700552},
quantum cryptography \citep{13-Groblacher_2006,15-Sit:17}, quantum
memory \citep{5-nicolas2014quantum}, chirality characterization of
crystals \citep{7-PhysRevB.91.094112}, holographic ghost imaging
\citep{26-PhysRevLett.103.083602}, spiral phase contrast imaging
\citep{25-Furhapter:05}, and particle control \citep{Prentice2004ManipulationAF},
among others. For many of these applications, state optimization processes
are crucial. Quantum weak measurement techniques \citep{Aharonov},
introduced by Aharonov, Albert, and Vaidman in the late 1980s, provide
a promising pathway for optimizing quantum states for specific tasks.

Researchers have confirmed the usefulness of quantum weak measurements
in enhancing the inherent properties of quantum states \citep{Y-2-Turek_2020,Y-Turek:2020gci,Y-PhysRevA.105.022210,Y-PhysRevA.105.022608,Y-TUREK2023128663,Y-Turek:2021qtx,Y-Turek:2022fuy,Turek_2020}.
Numerous theoretical and experimental studies have also explored the
application of orbital-angular-momentum (OAM) pointer states \citep{31-Dennis_2012,Turek_2015,333-PhysRevA.89.053816,333-PhysRevA.93.063841,333-zhu2019measuring,32-GOtte_2012,33-PhysRevLett.109.183903,34-Gotte:13,35-PhysRevLett.112.200401}.
Refs. \citep{333-PhysRevA.93.063841,333-PhysRevA.89.053816} introduced
weak measurement methods for determining the topological charge of
OAM beams and demonstrated that, for large orbital angular momentum
$l$, these methods are more efficient compared to earlier approaches
\citep{28-1-PhysRevLett.88.257901,28-4-Belmonte:12,28-2-PhysRevLett.105.153601,28-3-genevet2012holographic,29-PhysRevLett.111.074801,30-PhysRevA.89.025803,31-PhysRevA.89.053818}.
One of us, in collaboration, investigated high-order Laguerre-Gaussian
and Hermite-Gaussian (HG) mode pointers for postselected von Neumann
measurements \citep{Turek_2015}. They have found that high-order
LG and HG mode beams offer no advantage in precision measurements,
as characterized by the signal-to-noise ratio (SNR), compared to the
fundamental Gaussian beam. However, the superposition of high-order
LG and HG modes demonstrated advantages in precision measurements
due to interference effects between different modes. To our knowledge,
this proposal has not yet been further explored.

In applications involving OAM pointers, optical vortex beams often
undergo superposition due to imperfections in the generating apparatus
and the presence of noise. However, interference can sometimes enhance
the state quality and improve the resulting superposition. Therefore,
it is essential to investigate the impact of postselected von Neumann-type
states on the superposition of OAM beams.

In this paper, we have investigated the effects of von Neumann quantum
measurements on the superposition of the LG modes with $l=0$ and
$l=1$. We perform measurements on one mode of a superposition state
composed of two Laguerre-Gaussian (LG) beams, treating the spatial
and polarization degrees of freedom as the measured system and pointer,
respectively. The initial pointer state is a typical Gaussian state
whose Wigner function consistently takes positive values. Without
approximations, we determine the final state of the pointer and analyze
its associated properties, including quadrature squeezing, spatial
intensity distribution, second-order cross-correlation, and Wigner
functions. Our findings demonstrate that, following postselected measurements,
the properties of the initial pointer state undergo significant changes
when appropriate coupling strength parameters and anomalous weak values
of the measured system's observables are applied. Notably, the initial
Gaussian state transforms into a non-Gaussian state after the postselection
process. We anticipate that the scheme presented in this study could
be an effective optimization method for optical vortex beams, enhancing
the efficiency of various OAM-based applications.

The remainder of this paper is organized as follows. In Sec. II,
we describe the setup of our theoretical model. Sec. III discusses
the effects of postselected von Neumann measurements on the quadrature
squeezing of superpositions of Gaussian and Laguerre-Gaussian states,
demonstrating that postselected weak measurements can alter the squeezing
of the initial state. In Sec. IV, we analyze the spatial distribution
of the final pointer state. The influence of von Neumann measurements
on the second-order cross-correlation and the phase-space distribution
of the final state is explored in Sec. V. To evaluate the advantages
of the initial state in precision measurements based on postselected
von Neumann measurements, we study the signal-to-noise ratio (SNR)
for both postselected and non-postselected cases in Sec. VI. Finally,
in Sec. VII, we present our conclusions and provide an outlook on
future work. Throughout this paper, we adopt the natural unit system
with $\hbar=1$.

\section{Model setup\label{II}}

\begin{figure}
\begin{centering}
\includegraphics[width=8cm]{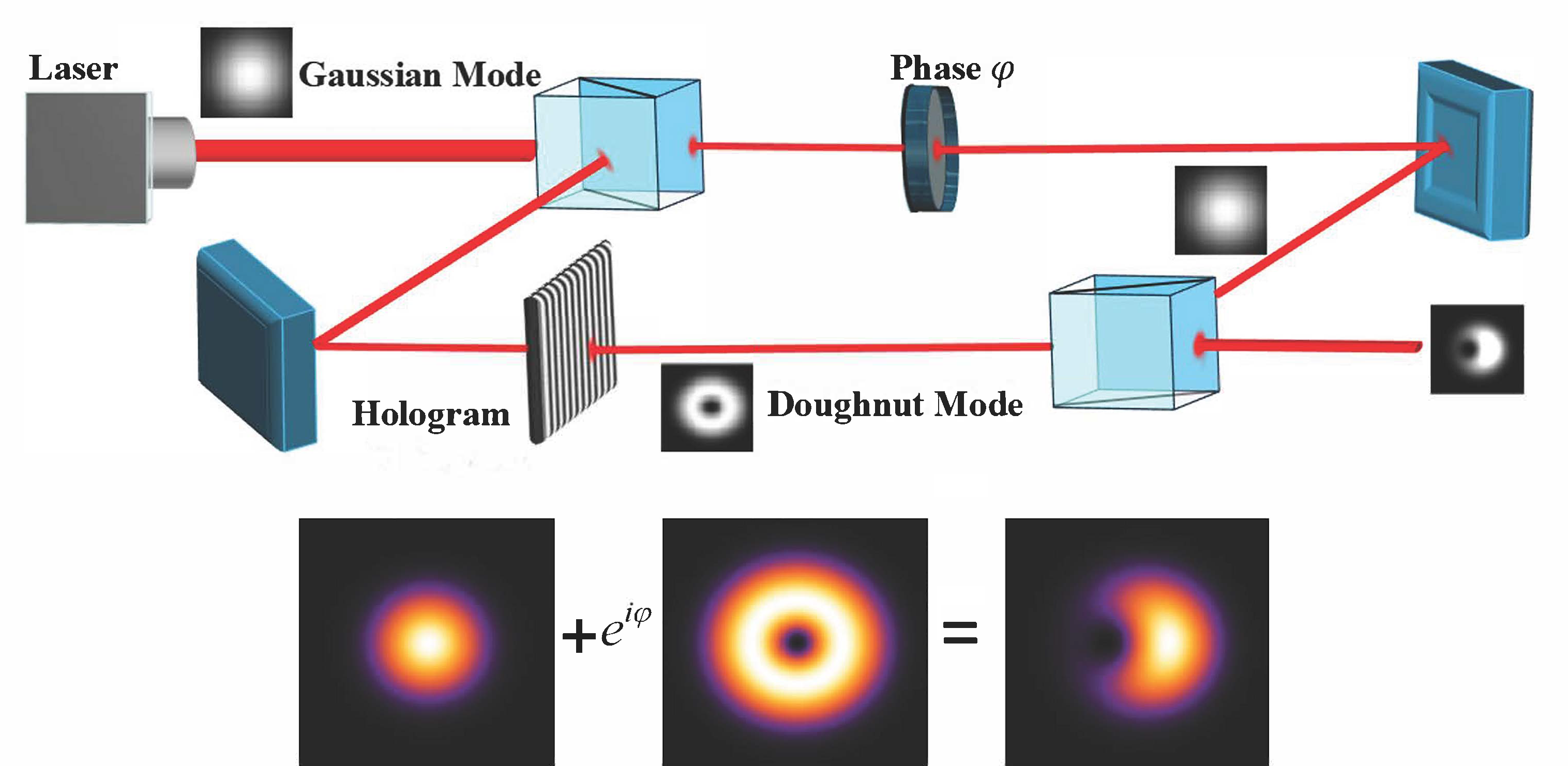}
\par\end{centering}
\caption{\label{fig:1}The setup for generating superpositions of the fundamental
Gaussian mode and the $LG_{01}$ mode, as investigated in Ref. \citep{Alipasha}}
\end{figure}

The superposition of various states plays a pivotal role in quantum
theory, a well-established concept with significant implications.
The superposition of higher-order OAM states also finds numerous applications
\citep{Alipasha,natureref}. Researchers have developed a variety
of experimental techniques to generate such superpositions \citep{Alipasha,PhysRevA.81.043844,2017A,Kotlyar:19,PhysRevA.103.063704,Wu:24}
. Among these, a straightforward method investigated in Ref. \citep{Alipasha}
involves using a Mach-Zehnder interferometer. In this approach, the
initial beam splitter divides the input beam into two paths. Each
optical path passes through a hologram, creating a superposition of
Gaussian and $LG_{01}$ modes. This process is achieved by placing
a hologram with a dislocation in one arm of the interferometer, as
shown in Fig. \ref{fig:1}(a). The method offers a distinct advantage:
by attenuating each arm and using phase plates, one can produce arbitrary
amplitude and relative phase superpositions without modifying the
experimental setup. The interference pattern resulting from the superposition
of the $LG_{00}$ mode and the $LG_{01}$ mode demonstrates this capability.
We can express its mathematical expression as: 
\begin{equation}
\left|\Psi_{i}\right\rangle =\frac{1}{\sqrt{1+\gamma^{2}}}\left[|\phi_{0}\rangle+\gamma\mathrm{e}^{\mathrm{i}\varphi}|\phi_{1}\rangle\right],\label{eq: ini}
\end{equation}
 where $\vert\phi_{l}\rangle=\int dxdy\phi_{l}(x,y)\vert x\rangle\vert y\rangle$
and the amplitude distribution of the $LG$ modes with radial indices
$p=0$ is characterized by 
\begin{equation}
\phi_{l}(x,y)=N\left[x+iy\ sgn(l)\right]^{\vert l\vert}\exp\left(-\frac{x^{2}+y^{2}}{2\sigma^{2}}\right).
\end{equation}
Here, $\sigma$ represents the variance for the case state with $l=0$,
sgn(.) is the sign function, and $N$ is a normalizing constant ensuring
that $\int dxdy\vert\phi_{l}(x,y)\vert^{2}=1$.

Meanwhile, the LG mode, which can be generated from the fundamental
Gaussian mode $\vert0,0\rangle$, is expressed in the following form
\citep{DELIMABERNARDO2014194}: 
\begin{equation}
\left|s,q\right\rangle _{LG}=\frac{1}{\sqrt{\alpha!\beta!}}\left(\hat{a}_{+}^{\dagger}\right)^{\alpha}\left(\hat{a}_{-}^{\dagger}\right)^{\beta}\left|0,0\right\rangle _{HG},
\end{equation}
 here, $\hat{a}_{\pm}=\left(\hat{a}\mp i\hat{b}\right)/\sqrt{2}$
and the integers $\alpha$ and $\beta$ are defined as $\alpha=(s+q)/2$
and $\beta=(s-q)/2$. The parameters $s$ and $q$ are related to
the radial and azimuthal indexes of the $LG$ modes by the relations
$l=q$ and $p=(s-\vert l\vert)/2$. Furthermore, $\hat{a}$ and $\hat{b}$
denote the annihilation operators corresponding to the $a$ and $b$
modes of the $HG$ beam. These operators act as follows: $\hat{a}\vert n,m\rangle=\hat{a}\vert n\rangle_{a}\vert m\rangle_{b}=\sqrt{n}\vert n-1,m\rangle$
and $\hat{b}\vert n,m\rangle=\hat{b}\vert n\rangle_{a}\vert m\rangle_{b}=\sqrt{m}\vert n,m-1\rangle$.
Using Eq. (\ref{eq: ini}), we can rewrite the state $\vert\Psi_{i}\rangle$
in terms of $HG$ modes as: 

\begin{align}
\vert\Psi_{i}\rangle & =\frac{1}{\sqrt{1+\gamma^{2}}}\left[\vert0,0\rangle_{HG}+\frac{\gamma\mathrm{e}^{\mathrm{i}\varphi}}{\sqrt{2}}\left(\vert1,0\rangle_{HG}+i\vert0,1\rangle_{HG}\right)\right]\nonumber \\
 & =\frac{1}{\sqrt{1+\gamma^{2}}}\left[\left(\vert0\rangle+\frac{\gamma\mathrm{e}^{\mathrm{i}\varphi}}{\sqrt{2}}\vert1\rangle\right)\vert0\rangle+i\frac{\gamma\mathrm{e}^{\mathrm{i}\varphi}}{\sqrt{2}}\vert0\rangle\vert1\rangle\right].\label{eq:4-1}
\end{align}
Here, the $HG$ modes defined as 
\begin{equation}
\vert n,m\rangle=\frac{1}{\sqrt{m!n!}}\left(\hat{a}\right)^{m}\left(\hat{b}\right)^{n}\left|0,0\right\rangle _{HG}.
\end{equation}
This is an entanglement state between two modes of the $HG$ beam.
In this study, we explore the properties of this state after a postselected
measurement. We use the spatial and polarization degrees of freedom
of $\vert\Psi_{i}\rangle$ as the measuring system (pointer) and the
measured system, respectively.

\begin{figure}
\begin{centering}
\includegraphics[width=8cm]{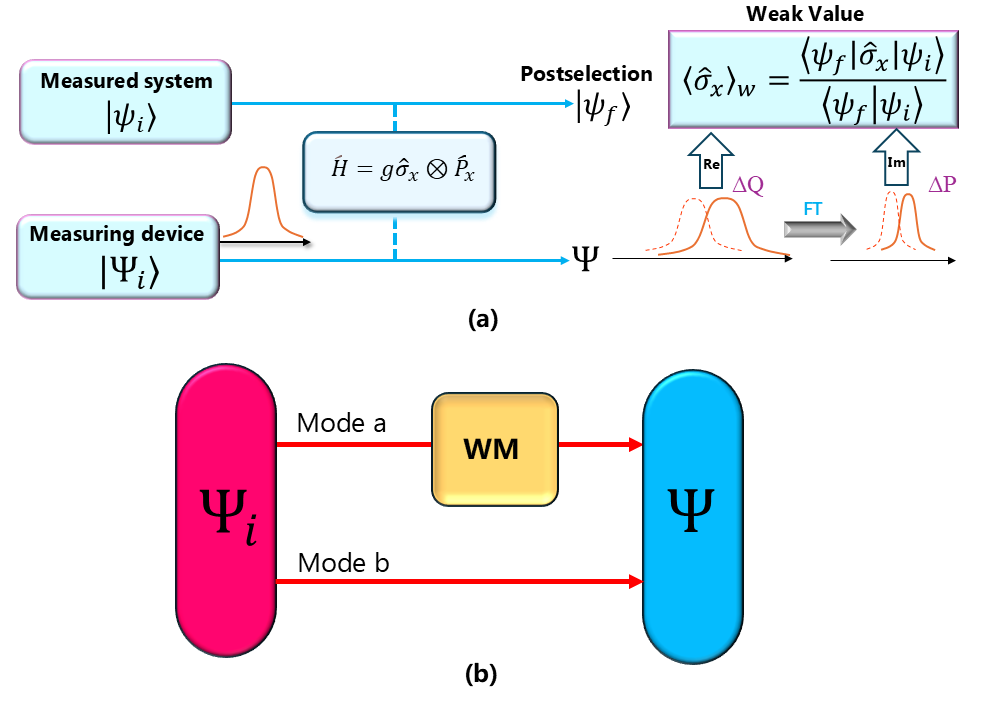}
\par\end{centering}
\caption{(a) Schematic diagram of the weak measurement (WM) process. The standard
procedure of WM involves four key steps: (i) The initial state of
the measured system is prepared as $\vert\psi_{i}\rangle$, and the
measuring device is initialized in the state $\vert\phi_{i}\rangle$.
(ii) A weak interaction occurs between the measured system and the
measuring device, causing the composite system to evolve. In this
model, the weak interaction is described by the Hamiltonian $H=g\,\hat{\sigma}_{x}\otimes\hat{P}_{x}$,
where the coupling constant $g$ characterizes the bilinear coupling.
(iii) After some evolution, the entire system is projected onto the
postselected state $\vert\psi_{f}\rangle$ of the measured system.
This postselection extracts the desired values of the system's observable
by choosing a specific subensemble of samples before the final measurement.
(iv) The measurement result is determined by analyzing the shifts
in the pointer. In the postselected weak measurement process, the
observable value expressed as a function of the weak value has its
real part ($\Re$) and imaginary part ($\Im$) extracted from the
shifts in position and momentum of the measuring device, respectively.
The Fourier transform (FT) is used to convert position space into
momentum space. (b) Schematic setup for preparing the state $\vert\Psi\rangle$
via a postselected von Neumann measurement.}

\end{figure}

For simplicity, in this work, we consider only a measurement performed
on one mode of the system. According to the standard measurement theory
proposed by von Neumann \citep{vonNeumann+2018}, the coupling between
the measured system and the pointer is: 

\begin{align}
H & =g\text{ \ensuremath{\hat{\sigma}_{x}}\ensuremath{\otimes}\ensuremath{\hat{P_{x}}}.}\label{eq:1}
\end{align}
Here, the $g$ represents the interaction coupling parameter between
the pointer and the measured system, and $\hat{\sigma}_{x}=\vert D\rangle\langle D\vert-\vert A\rangle\langle A\vert$
is the operator of the measured system, where $\vert D\rangle=\frac{1}{\sqrt{2}}\left(\vert H\rangle+\vert V\rangle\right)$
and $\vert A\rangle=\frac{1}{\sqrt{2}}\left(\vert H\rangle-\vert V\rangle\right)$
represent the diagonal and anti-diagonal polarizations in the horizontal
$\vert H\rangle$ and vertical $\vert V\rangle$ polarization bases
of the beam, respectively. The operator $\hat{P}_{x}$ denotes the
momentum operator of the $a$ mode of the pointer, which is conjugate
to $\hat{X}=\int dx\ x\vert x\rangle\langle x\vert$, i.e., $\left[\hat{X},\hat{P}_{x}\right]=i$.
Assuming that the initial state of the measured system and the pointer
are $\vert\psi_{i}\rangle$ and $\vert\Psi_{i}\rangle$, respectively,
the state $\vert\psi_{i}\rangle\otimes\vert\Psi_{i}\rangle$ evolves
under the unitary evolution operator $\hat{U}(t)=\exp\left(-i\int_{0}^{t}\hat{H}d\tau\right)$
to:

\begin{align*}
\vert\Phi\rangle & =\exp\left(-i\int_{0}^{t}\hat{H}d\tau\right)\vert\psi_{i}\rangle\otimes\vert\Psi_{i}\rangle
\end{align*}

\begin{align}
 & =\frac{1}{2}\left[(\mathbb{I}+\hat{\sigma}_{x})\otimes\hat{D}\left(\frac{\Gamma}{2}\right)+(\mathbb{I}-\hat{\sigma}_{x})\otimes\hat{D}\left(-\frac{\Gamma}{2}\right)\right]\nonumber \\
 & \qquad\times\vert\psi_{i}\rangle\otimes\vert\Psi_{i}\rangle.\label{eq:3-1}
\end{align}
Here, $\hat{D}(\Gamma/2)=e^{\Gamma/2(\hat{a}^{\dagger}-\hat{a})}$
is the displacement operator, $\mathbb{I}$ is the $2\times2$ identity
matrix operator, and $\Gamma=\frac{gt}{\sigma}$ is the coupling strength
parameter. In the derivation of the above expression, we write the
momentum operator $\hat{P}$ in terms of the annihilation and creation
operators, $\hat{a}$ and $\hat{a}^{\dagger}$, as $\hat{P}=\frac{i}{2\sigma}\left(\hat{a}^{\dagger}-\hat{a}\right)$,
where $\sigma$ is the size of the fundamental Gaussian beam. The
parameter $s$ is dimensionless and can take continuous values, characterizing
the measurement strength. If $0<\Gamma\ll1$, the measurement is weak,
whereas $\Gamma\gg1$, is classified as strong. The value of $s$
can be controlled experimentally in three ways, corresponding to adjustments
in $g$, $t$, and $\sigma$. Experimental research \citep{Nature2020}
has shown that the simplest and most direct way to adjust the coupling
strength parameter $s$ is by tuning the coupling duration $t$. Below,
we assume that the change in $\Gamma$ comes from $t$, while $g$
and $\sigma$ are fixed. For the implementation of the postselected
von Neumann measurement, the postselected state $\vert\psi_{f}\rangle$
is taken as in Eq. (\ref{eq:3-1}), and we can express the normalized
final state of the pointer as: 

\begin{equation}
\vert\Psi\rangle=\lambda\left[\left(1+\langle\hat{\sigma}_{x}\rangle_{w}\right)\hat{D}\left(\frac{\Gamma}{2}\right)+\left(1-\langle\hat{\sigma}_{x}\rangle_{w}\right)\hat{D}^{\dagger}\left(\frac{\Gamma}{2}\right)\right]\vert\Psi_{i}\rangle,\label{eq:4}
\end{equation}
where the normalization coefficient $\lambda$ is defined as 
\begin{equation}
\lambda=\left\{ \frac{1}{2}\left[1+\vert\langle\hat{\sigma}_{x}\rangle_{w}\vert^{2}+\left(1-\vert\langle\hat{\sigma}_{x}\rangle_{w}\vert^{2}\right)Re\text{\ensuremath{[I_{1}]}}\right]\right\} ^{-\frac{1}{2}}
\end{equation}
with 
\begin{equation}
I_{1}=\left\{ 1-\frac{1}{1+\gamma^{2}}\left[i\sqrt{2}\Gamma\gamma\sin(\varphi)+\frac{\left(\gamma\Gamma\right)^{2}}{2}\right]\right\} e^{-\frac{\Gamma^{2}}{2}}.
\end{equation}
In Eq. (\ref{eq:4}), the $\langle\hat{\sigma}_{x}\rangle_{w}$ is
the weak value of the measured operator $\hat{\sigma}_{x}$. In this
work, we assume the preselected and postselected states of the measured
system are given by $\vert\psi_{i}\rangle=\cos\text{\ensuremath{\left(\text{\ensuremath{\frac{\alpha}{2}}}\right)}}\vert H\rangle+e^{i\delta}\sin\left(\frac{\alpha}{2}\right)\vert V\rangle$
and $\vert\psi_{f}\rangle=\vert H\rangle$, respectively. For these
states, the weak value $\langle\hat{\sigma}_{x}\rangle_{w}$ is given
by:

\begin{equation}
\langle\hat{\sigma}_{x}\rangle_{w}=\frac{\langle\psi_{f}\vert\hat{\sigma}_{x}\vert\psi_{i}\rangle}{\langle\psi_{f}\vert\psi_{i}\rangle}=e^{i\delta}\tan\frac{\alpha}{2}.\label{eq:weak values}
\end{equation}
Here, $\delta\in[0,2\pi]$ and $\alpha\in[0,\pi)$. During the derivation
of $\vert\Psi\rangle$, we used the mathematical definition of the
displaced Fock state: 
\begin{equation}
\hat{D}\left(\alpha\right)\left|n\right\rangle =e^{-\frac{\vert\alpha\vert^{2}}{2}}\sum_{k=0}^{\infty}\left(\frac{n!}{k!}\right)^{\frac{1}{2}}\left(\alpha\right)^{k-n}L_{n}^{\left(k-n\right)}(\vert\alpha\vert^{2})\left|k\right\rangle ,\label{eq:11}
\end{equation}
where $L_{n}^{(k-n)}(.)$ are the generalized Laguerre polynomials.
Equation (\ref{eq:4}) represents the final state of the pointer after
the postselected von Neumann measurement. The weak value above can
exceed the normal range of observable values for $\sigma_{x}$ when
the pre- and postselected states are nearly orthogonal, and it can
even take complex values in the case where $\delta\neq0$. As mentioned
in the introduction, anomalously large weak values amplify tiny system
information and can also be used to optimize quantum states. Next,
we examine the effects of anomalous weak values of the measured system
observable on the intrinsic properties of $\vert\Psi\rangle$.

\section{\label{III}Effects on Quadrature squeezing}

In this subsection, we study the effects of postselected von Neumann
measurement on the quadrature squeezing of $\vert\Psi\rangle$. To
discuss the squeezing phenomenon, we define the two-mode quadrature
operators as \citep{Adesso2014ContinuousVQ}: 

\begin{align}
\hat{F}_{1} & =\frac{1}{2^{3/2}}(\hat{a}+\hat{b}+\hat{a}^{\dagger}+\hat{b}^{\dagger}),\label{eq:SQ-1}\\
\hat{F}_{2} & =\frac{1}{2^{3/2}i}(\hat{a}+\hat{b}-\hat{a}^{\dagger}-\hat{b}^{\dagger}).\label{eq:SQ-2}
\end{align}
One can verify that these two operators satisfies the commutation
relation, $\left[\hat{F}_{1},\hat{F}_{2}\right]=\frac{i}{2}$, and
the corresponding uncertainty relation for their fluctuations is

\begin{align}
\Delta\hat{F}_{1}^{2}\Delta\hat{F}_{2}^{2} & \geq\frac{1}{16},\label{eq:14}
\end{align}
where $\Delta F_{i}^{2}=\langle F_{i}^{2}\rangle-\langle F_{i}\rangle^{2}\,(i=1,2)$.
The squeezing parameter, which characterizes the quadrature squeezing
of the $i$-th component of $\vert\Psi\rangle$ (superpositions of
Gaussian and Laguerre-Gaussian states), is defined as follows: 

\begin{align}
Q_{i} & =\Delta\hat{F}_{i}^{2}-\frac{1}{4}.\label{eq:15}
\end{align}
As we can see, the values of $Q_{i}$ are bounded by $Q_{i}\ge-\frac{1}{4}$,
and the $i$-th component of the quadrature operators of $\vert\Psi\rangle$
is considered squeezed if $-\frac{1}{4}\le Q_{i}<0$. After some algebra,
we can obtain the squeezing parameters $Q_{i}$ for the final state
$\vert\Psi\rangle$ as: 

\begin{align}
Q_{1,\Psi} & =\frac{1}{4}\left[\langle\hat{a}^{\dagger}\hat{a}\rangle+\langle\hat{b}^{\dagger}\hat{b}\rangle+\langle\hat{a}^{\dagger}\hat{b}\rangle+\langle\hat{a}\hat{b}^{\dagger}\rangle+\langle\hat{a}\hat{b}+\hat{a}^{\dagger}\hat{b}^{\dagger}\rangle\right]\nonumber \\
 & \qquad+\frac{1}{8}\left[\langle\hat{a}^{2}+\hat{a}^{\dagger2}\rangle+\langle\hat{b}^{2}+\hat{b}^{\dagger2}\rangle\right]\nonumber \\
 & \qquad-\frac{1}{8}\left[\langle\hat{a}+\hat{a}^{\dagger}\rangle+\langle\hat{b}+\hat{b}^{\dagger}\rangle\right]^{2}.\label{eq:7}
\end{align}

\begin{align}
Q_{2,\Psi} & =\frac{1}{4}\left[\langle\hat{a}^{\dagger}\hat{a}\rangle+\langle\hat{b}^{\dagger}\hat{b}\rangle+\langle\hat{a}^{\dagger}\hat{b}\rangle+\langle\hat{a}\hat{b}^{\dagger}\rangle+\langle\hat{a}\hat{b}+\hat{a}^{\dagger}\hat{b}^{\dagger}\rangle\right]\nonumber \\
 & \qquad-\frac{1}{8}\left[\langle\hat{a}^{2}+\hat{a}^{\dagger2}\rangle+\langle\hat{b}^{2}+\hat{b}^{\dagger2}\rangle\right]\nonumber \\
 & \qquad+\frac{1}{8}\left[\langle\hat{a}+\hat{a}^{\dagger}\rangle+\langle\hat{b}+\hat{b}^{\dagger}\rangle\right]^{2}.\label{eq:18}
\end{align}
Here, $\langle\cdots\rangle$ denotes the average values of the associated
operators under the state $\vert\Psi\rangle$, and Appendix \ref{sec:A1}
provides the analytic expressions for these expectation values.

\begin{figure}
\begin{centering}
\includegraphics[width=8cm]{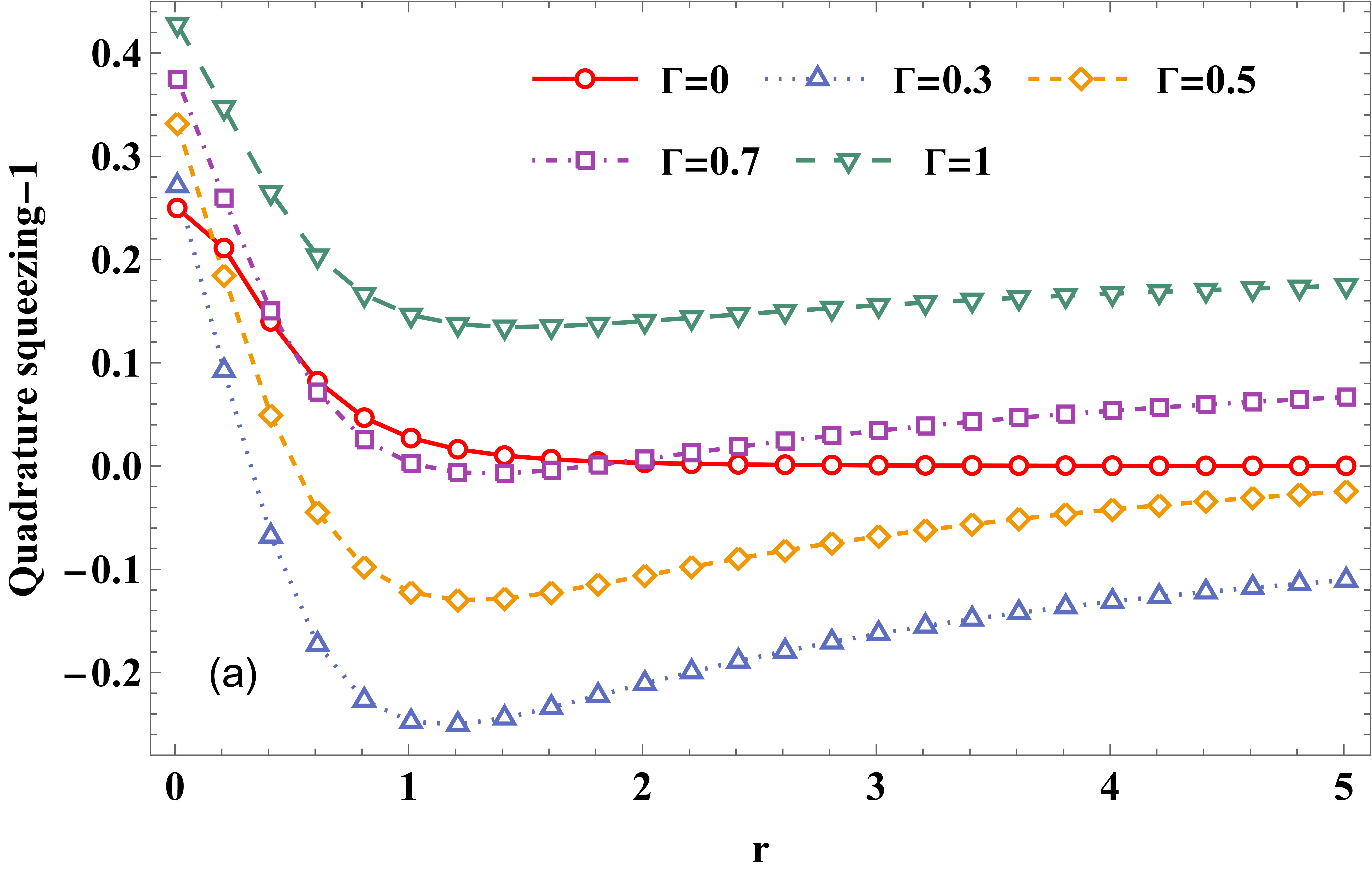}
\par\end{centering}
\begin{centering}
\includegraphics[width=8cm]{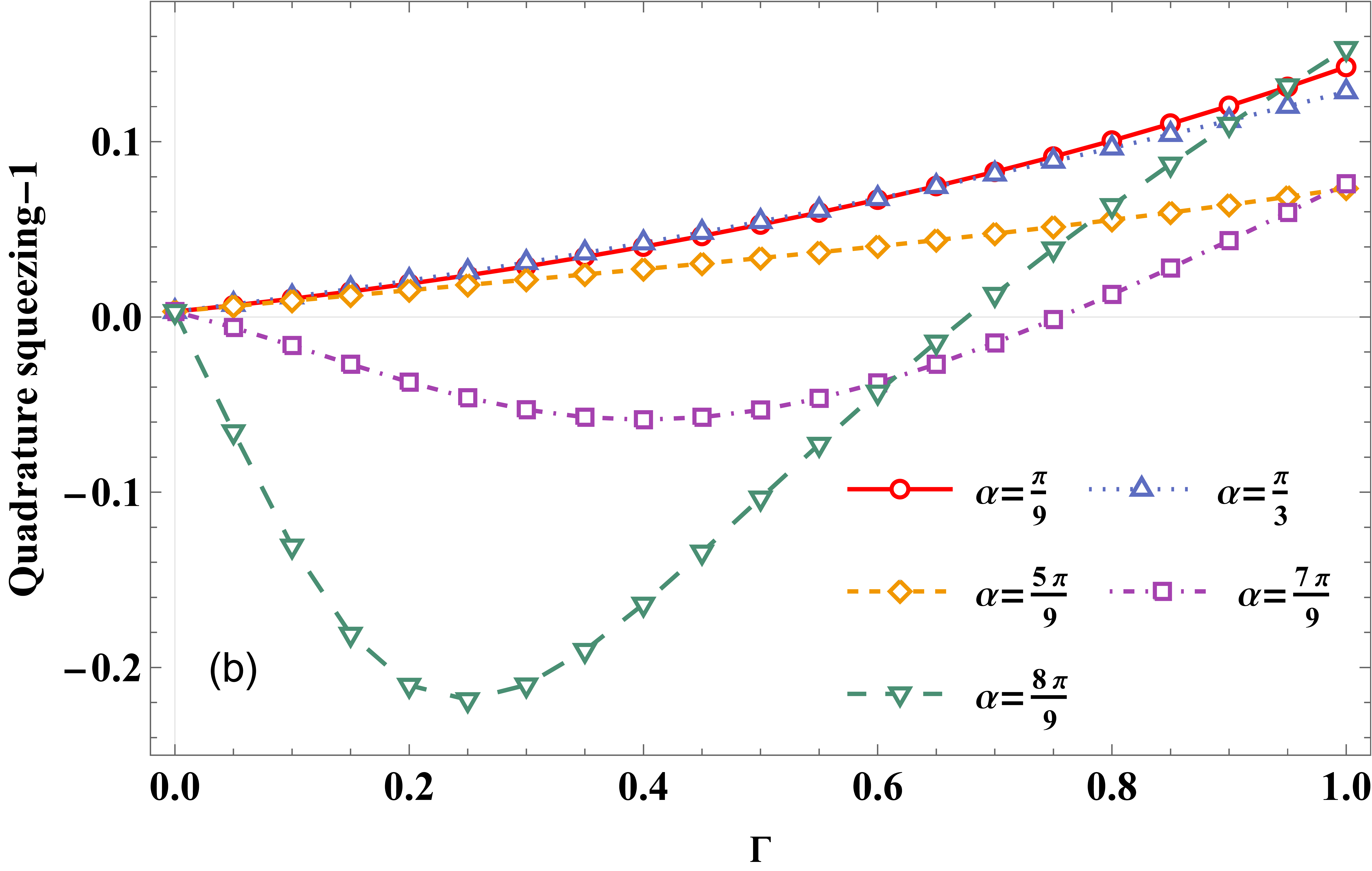}
\par\end{centering}
\begin{centering}
\includegraphics[width=8cm]{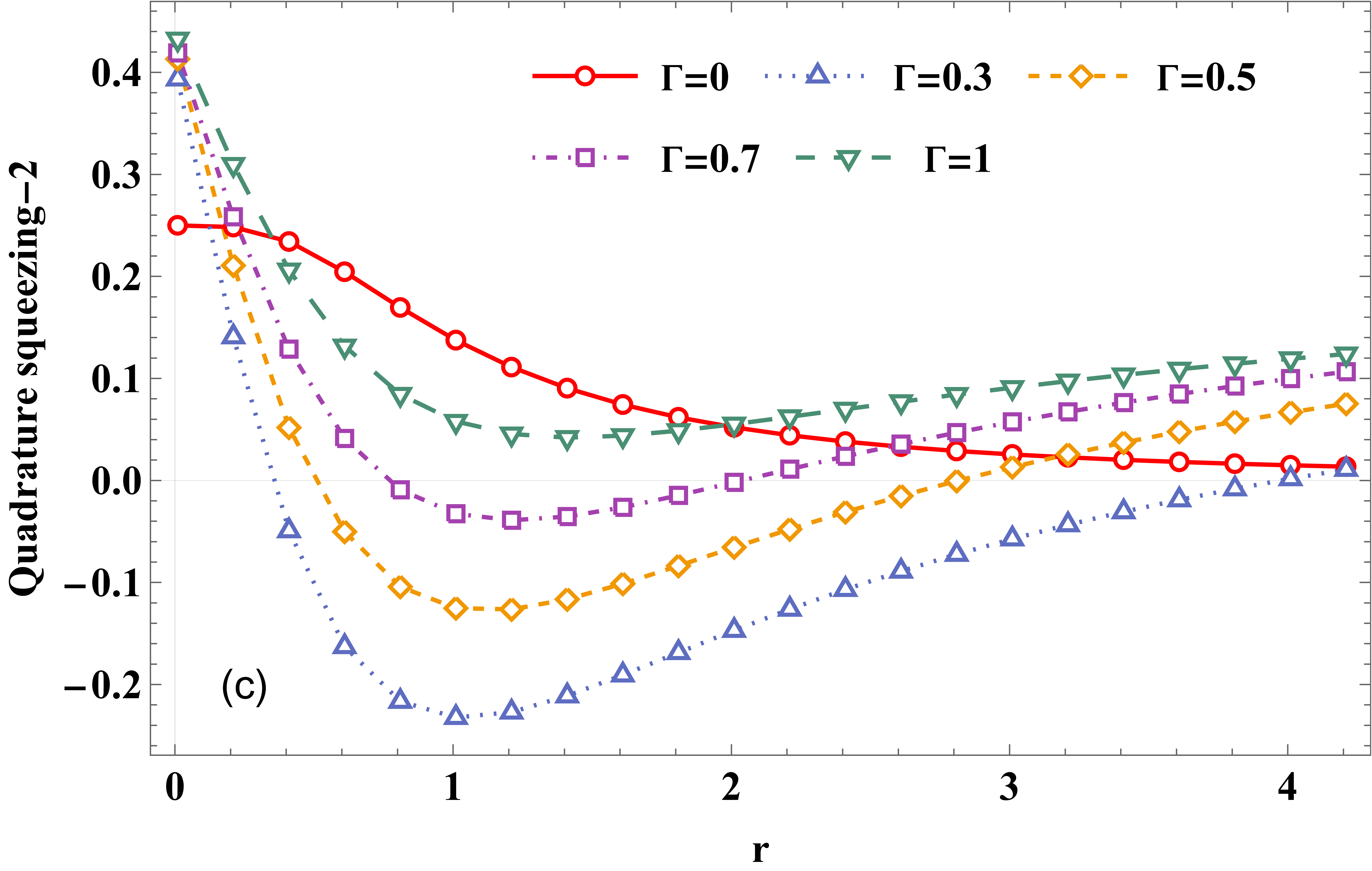}
\par\end{centering}
\begin{centering}
\includegraphics[width=8cm]{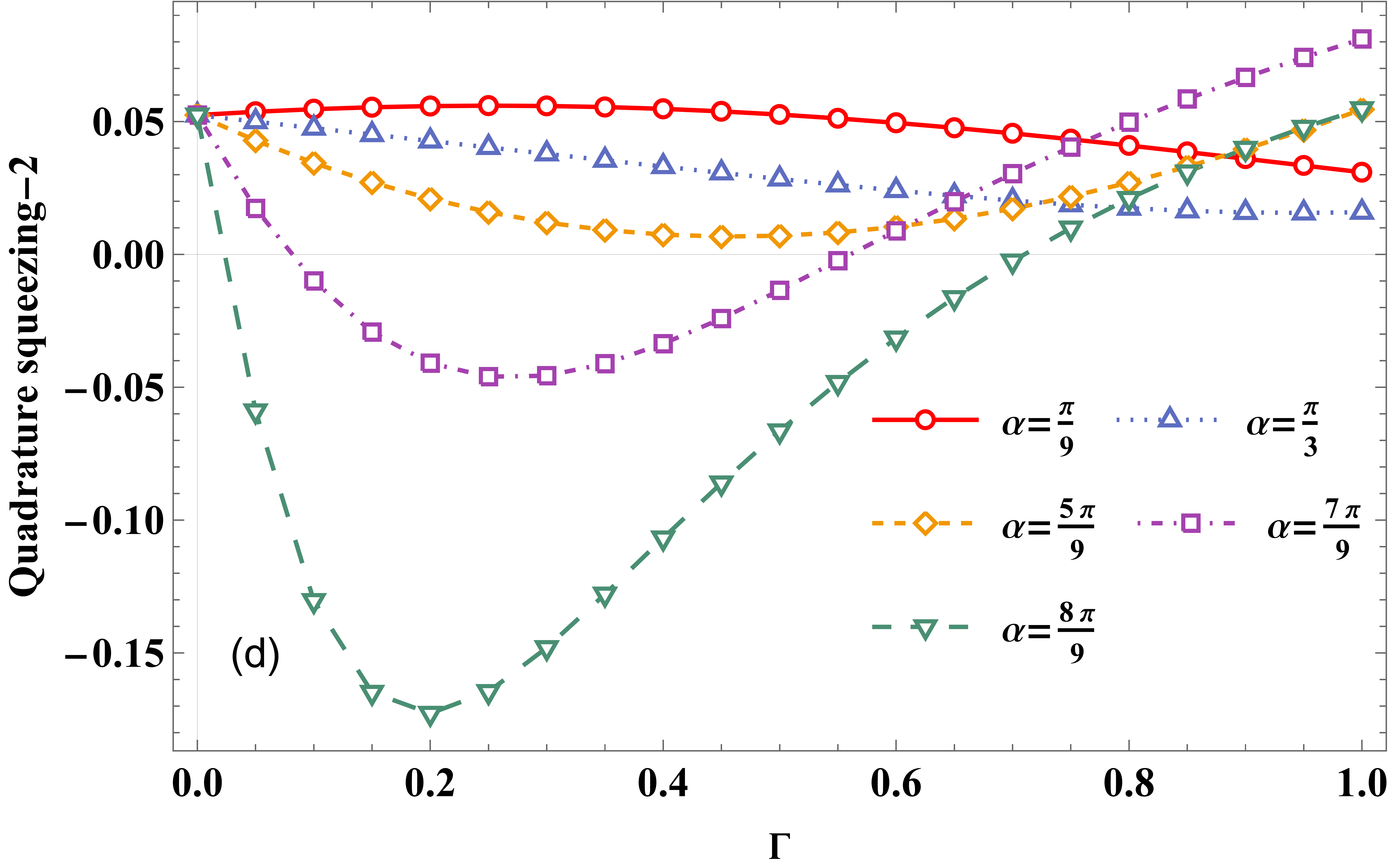}
\par\end{centering}
\caption{\label{fig:3}Quadratic squeezing parameters $Q_{1}$ and $Q_{2}$.
(a) $Q_{1}$ as a function of $r$ for different coupling strengths
$\Gamma$, with the weak value fixed at $\alpha=\frac{8\pi}{9}$,
(b) $Q_{1}$ as a function of the weak value with coupling $\Gamma$
at $r=2$. (c) $Q_{2}$ as a function of $r$ for different coupling
strengths $\Gamma$, with the weak value fixed at $\alpha=\frac{8\pi}{9}$.
(d) $Q_{2}$ as a function of thee weak values characterized by $\Gamma$
at $r=2$. Here, $\delta=0$ and $\varphi=\frac{\pi}{2}$.}
\end{figure}

To clearly explain the effects of the postselected von Neumann measurement
on the quadrature squeezing of $\vert\Psi\rangle$, we rely on numerical
analysis and the corresponding results shown in Fig. \ref{fig:3}.
In Figs. \ref{fig:3}(a) and (c), we plot $Q_{1}$ and $Q_{2}$ as
functions of $r=\sqrt{x^{2}+y^{2}}$ for different coupling strength
parameters $\Gamma$, while fixing the weak value parameter at $\alpha=8\pi/9$.
Initially, the state $\vert\Psi_{i}\rangle$ exhibits no quadrature
squeezing effect across all parameter regions. Additionally, for large
anomalous weak values, such as $\langle\sigma_{x}\rangle_{w}=5.671$
($\alpha=8\pi/9$), the resulting squeezing effect is greater than
that of the original state ($\Gamma=0$) in the weak measurement regime.
This result reflects the amplification effect of the weak value. Figs.
\ref{fig:3}(b) and (d) present the changes in $Q_{1}$ and $Q_{2}$
as functions of the coupling strength parameter $\Gamma$ for different
weak values while fixing $r=2$. In the appropriate weak measurement
regime (small $\Gamma$), the squeezing of $\vert\Psi\rangle$ increases
with the weak value. We investigate the effects of the postselected
von Neumann measurement on the properties of the output state $\vert\Psi\rangle$.

\section{\label{IV}Effects on intensity distribution }

Postselected weak measurements on the measured system can alter the
inherent characteristics of the pointer due to weak value amplification.
This effect has also been confirmed in OAM-based pointer measurement
problems. In this section, we investigate the impact of a postselected
von Neumann measurement on the intensity distribution of the state,
defined by superpositions of LG modes, such as in Eq. (\ref{eq:3-1}).
We can rewrite the initial state $\vert\Psi_{i}\rangle$ in the coordinate
representation as: 
\begin{align}
\Psi_{i}(x,y) & =\langle x,y\vert\Psi_{i}\rangle\nonumber \\
 & =\frac{1}{\sqrt{1+\gamma^{2}}}\left[\phi_{0}(x,y)+\frac{\gamma e^{i\varphi}}{\sqrt{2}}\frac{1}{\sigma}\left(x+iy\right)\phi_{0}(x,y)\right]
\end{align}
 where 
\[
\phi_{0}(x,y)=\left(\frac{1}{\pi\sigma^{2}}\right)^{\frac{1}{2}}e^{-\frac{x^{2}+y^{2}}{2\sigma^{2}}}
\]
 This state is the two-dimensional Gaussian profile corresponding
to the fundamental mode of the HG and LG states. As introduced in
Sec. \ref{II}, after the postselected measurement, the state of the
pointer is changed to $\vert\Psi\rangle$ (see Eq. (\ref{eq:4})),
and its expression in the coordinate representation is given by:
\begin{align}
\Psi(x,y) & =\frac{\kappa}{\sqrt{1+\gamma^{2}}}\{t_{-}M_{-s}+t_{+}M_{+s}+t_{-}T_{-}+t_{+}T_{+}\nonumber \\
 & +t_{-}K_{-}+t_{+}K_{+}\}\label{eq:20}
\end{align}
 where $t_{\pm}=1\pm\langle\sigma_{x}\rangle_{w}$, $s=\Gamma/2$,
$M_{\pm s}=\phi_{\pm s}(x)\psi(y)$, $K_{\pm}=i\frac{\sqrt{2}y}{\sigma}\gamma e^{i\varphi}\phi_{\pm s}(x)\psi(y),$
and $T_{\pm}=\frac{\gamma e^{i\varphi}}{\sqrt{2}}\left[\pm(1-\sqrt{2})s+\frac{2x}{\sigma}\right]\phi_{\pm s}(x)\psi(y)$
with 
\begin{equation}
\phi_{s}(x)=\left(\frac{1}{\pi\sigma^{2}}\right)^{\frac{1}{4}}e^{-\frac{s^{2}}{2}}e^{\frac{x^{2}}{2\sigma^{2}}}e^{-(\frac{x}{\sigma}-\frac{s}{\sqrt{2}})^{2}},\label{eq:21}
\end{equation}
and 
\begin{equation}
\psi(y)=\left(\frac{1}{\pi\sigma^{2}}\right)^{\frac{1}{4}}e^{-\frac{y^{2}}{2\sigma^{2}}}.\label{eq:22}
\end{equation}
 
\begin{flushleft}
\begin{figure}
\begin{centering}
\includegraphics[width=8cm,totalheight=14cm]{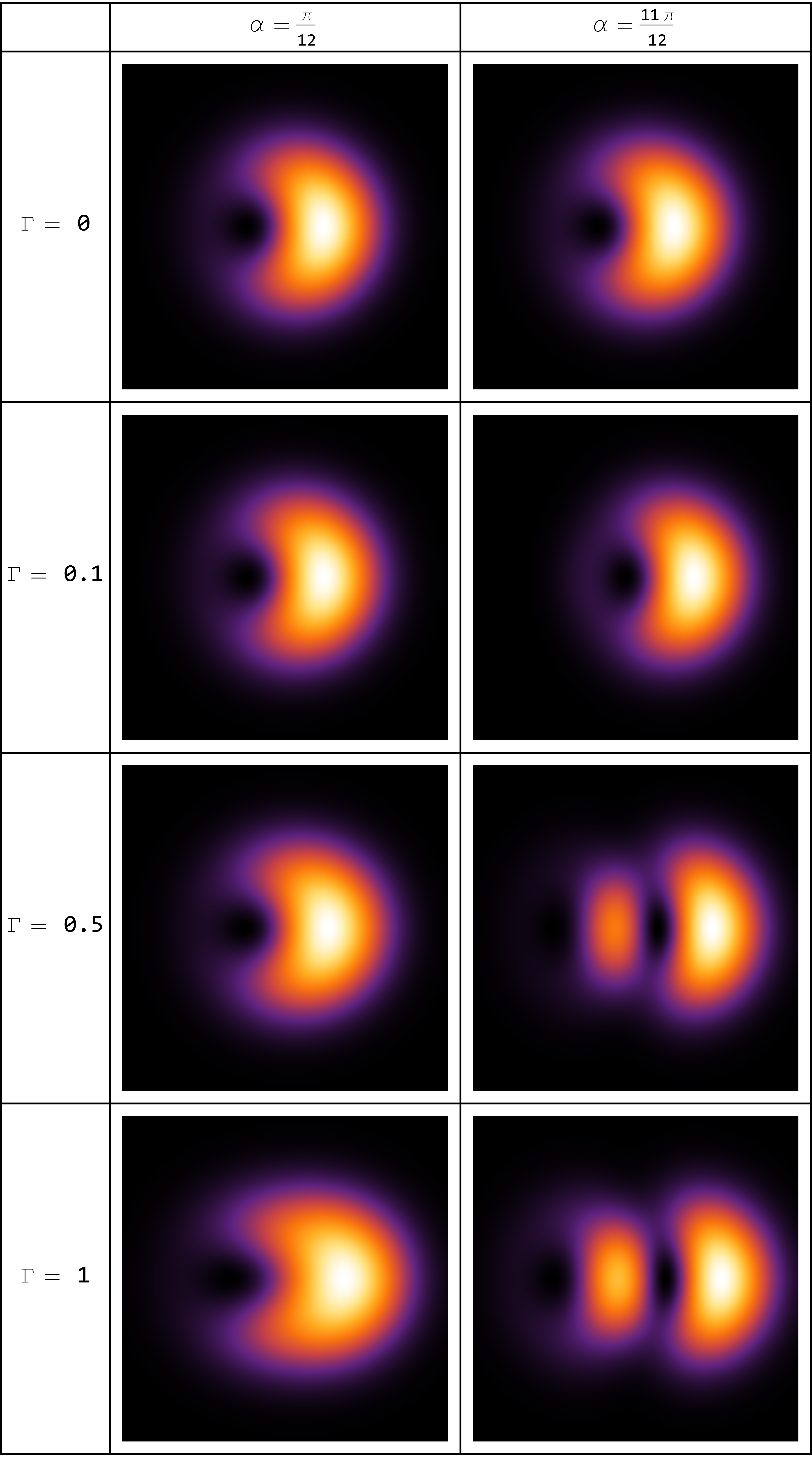}
\par\end{centering}
\caption{Intensity distribution after the postselected measurement for different
measurement coupling strength parameter $\Gamma$ and weak value parameter
$\alpha$. Here we take $\varphi=0$,$\delta=0$,$\sigma=1$ and $\gamma=1$.\label{fig:2}}
\end{figure}
\par\end{flushleft}

To examine the effects of postselected von Neumann measurement on
the spatial intensity profile of the superposition of LG modes defined
in Eq. (\ref{eq: ini}), we plot the intensity distribution of the
state $\vert\Psi\rangle$ as shown in Fig. \ref{fig:2}. For comparison,
in Fig. \ref{fig:2} , we present the intensity distribution of the
state $\vert\Psi\rangle$ for small and large weak values, as for
different coupling strengths $\Gamma$. As indicated in the first
column of Fig. \ref{fig:2}, for small weak values, i.e., $\langle\sigma_{x}\rangle_{w}=0.132$,
the spatial profile of $\vert\Psi\rangle$ does not change dramatically
and retains its initial shape as the coupling strength parameter $\Gamma$
increases. As shown in the second column of Fig. \ref{fig:2}, after
a postselected measurement with a large weak value, i.e., $\langle\sigma_{x}\rangle_{w}=7.596$,
the spatial intensity distribution of the initial state $\vert\Psi_{i}\rangle$
changes significantly as $\Gamma$ increases. In particular, for large
weak values and $\Gamma=1$, the intensity distribution of $\vert\Psi\rangle$
separates into two parts, with each part exhibiting a zero-intensity
region at the edges of the images, similar to the initial case. From
the above analysis, we can confirm that the spatial intensity distribution
of the OAM state notably changes when considering large weak values
of the measured observable after a postselected von Neumann measurement.
This intriguing result also suggests the potential applicability of
postselected von Neumann measurements, characterized by weak values,
in OAM-based state engineering processes.

\section{Second-order cross-correlation function and phase space distribution
\label{V}}

To further investigate the effects of postselected von Neumann measurement
on the properties of $\vert\Psi\rangle$, in this section, we study
the quantum statistics and phase space distribution of $\vert\Psi\rangle$
for different system parameters. 

\subsection{\label{subsec:4-1}Second-order cross-correlation function }

In this subsection, we study the second-order cross-correlation function
$g_{a,b}^{(2)}$ of $\vert\Psi\rangle$. The second-order cross-correlation
function of a two-mode radiation field is defined as \citep{REN2019106,GIRI2017140}: 

\begin{equation}
g_{a,b,\Psi}^{(2)}=\frac{\langle\hat{a}^{\dagger}\hat{a}\hat{b}^{\dagger}\hat{b}\rangle}{\langle\hat{a}^{\dagger}\hat{a}\rangle\langle\hat{b}^{\dagger}\hat{b}\rangle},
\end{equation}
here, $\langle\hat{a}^{\dagger}\hat{a}\hat{b}^{\dagger}\hat{b}\rangle$
represents the intensity-intensity correlation between the two modes,
and $\langle\hat{a}^{\dagger}\hat{a}\rangle$ and $\langle\hat{b}^{\dagger}\hat{b}\rangle$
denote the mean photon numbers for each mode, respectively. This correlation
characterizes the relationship between photons in different modes.
If $g_{a,b,\Psi}^{(2)}>1$, there exists a correlation between the
$a$-mode and $b$-mode of the two-mode radiation field; otherwise,
the modes are inversely correlated. To analyze the properties of $g_{a,b,\Psi}^{(2)}$,
we first derive the average values of $\langle\hat{a}^{\dagger}\hat{a}\hat{b}^{\dagger}\hat{b}\rangle$,
$\langle\hat{a}^{\dagger}\hat{a}\rangle$, and $\langle\hat{b}^{\dagger}\hat{b}\rangle$
under the state $\vert\Psi\rangle$. Appendix \ref{sec:A1} lists
their explicit, cumbersome expressions.

To examine how the postselected von Neumann measurement affects the
SOCC of the state $\vert\Psi\rangle$, we plot $g_{a,b,\Psi}^{(2)}$
for different system parameters associated with the postselected von
Neumann measurement. Fig. \ref{fig:G2-X} presents the corresponding
numerical results. The case $\Gamma=0$ corresponds to no interaction
between the pointer and the measured system. Initially, we observe
no correlation between the two modes of $\vert\Psi\rangle$ (see the
red-colored curves of Fig. \ref{fig:G2-X}). Our numerical results
indicate that after the postselected measurement, no correlation occurs;
however, as the weak value increases, $g_{a,b,\Psi}^{(2)}$ approaches
one. This behavior is due to the signal amplification effects of anomalously
large weak values. 
\begin{center}
\begin{figure}
\begin{centering}
\includegraphics[width=8cm]{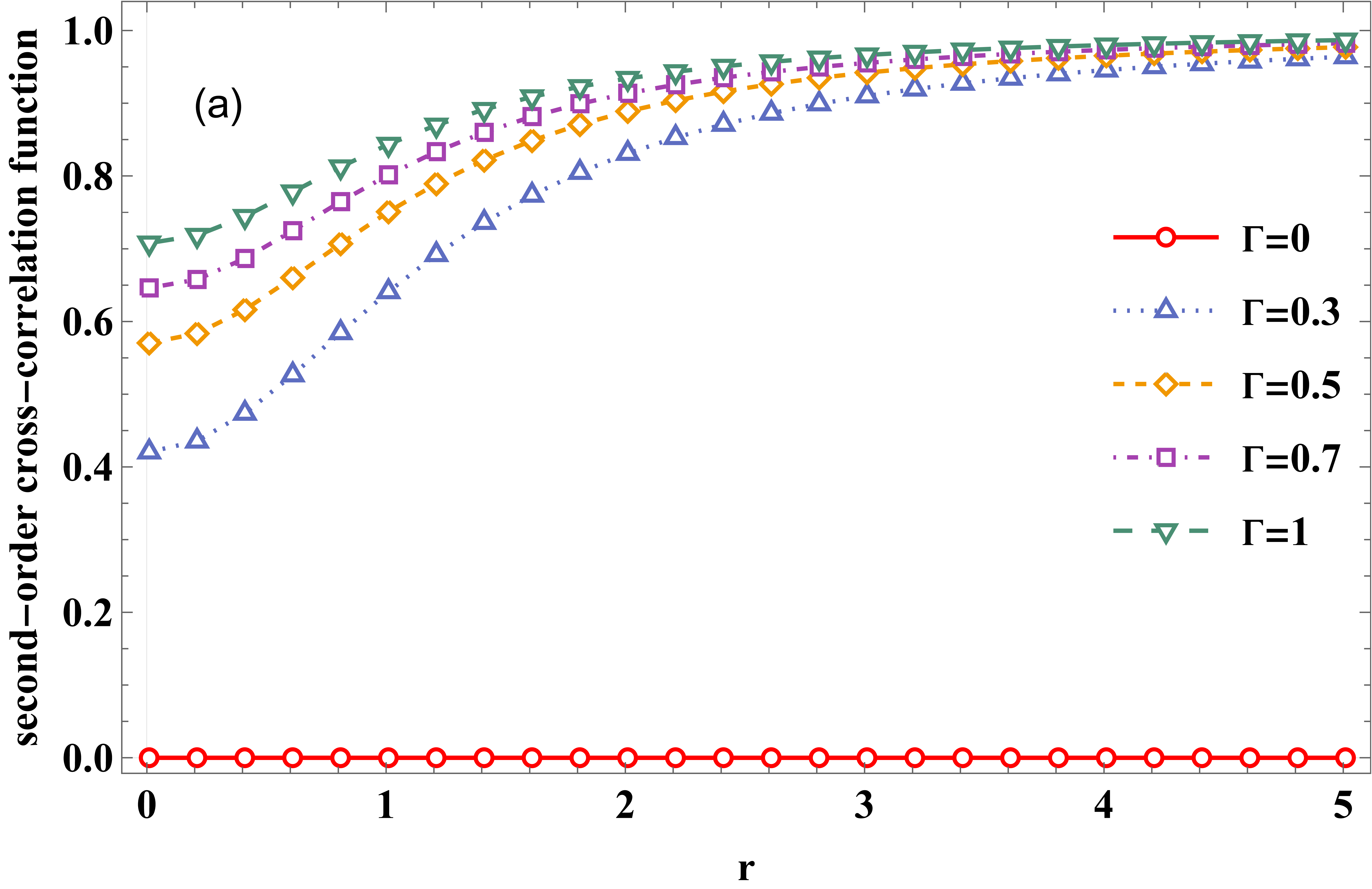}
\par\end{centering}
\begin{centering}
\includegraphics[width=8cm]{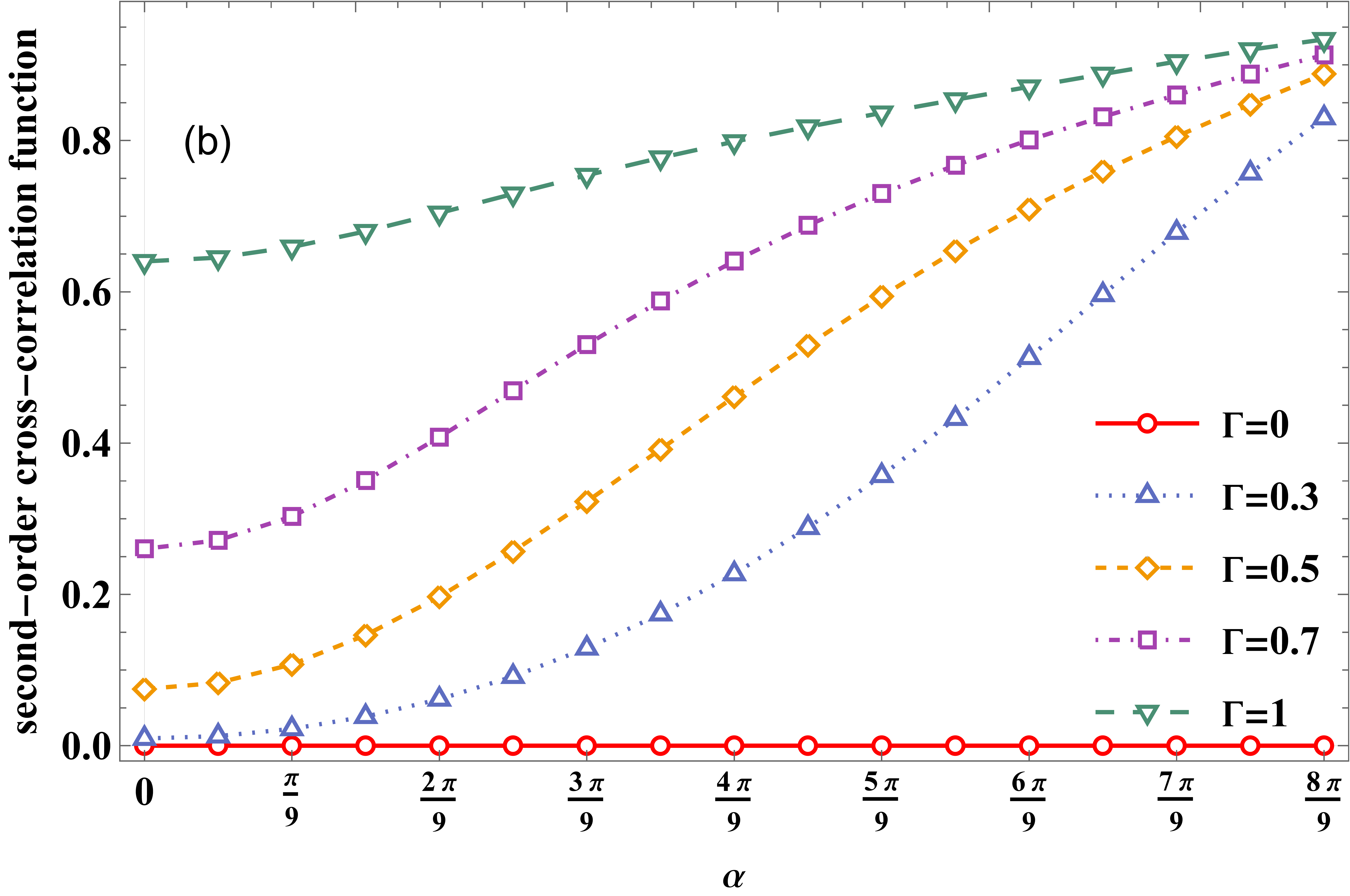}
\par\end{centering}
\caption{(a) Second-order cross-correlation as function of $r$ for different
coupling strengths $\Gamma$, with the weak value fixed at $\alpha=\frac{8\pi}{9}$.
(b) Second-order cross-correlation as function of the weak value $\alpha$
for different coupling strengths $\Gamma$, with $r=2$. Here, $\delta=0$
and $\varphi=\frac{\pi}{2}$.\label{fig:G2-X}}
\end{figure}
\par\end{center}

\subsection{\label{subsec:4-2}Wigner function}

To deeply understand the effects of postselected von Neumann measurements
on the properties of $\vert\Psi\rangle$, we examine the phase space
distribution by calculating its Wigner function. The general expression
for the Wigner function of a state $\rho=\vert\Psi\rangle\langle\Psi\vert$
is \citep{Int,Agarwal2013}: 

\begin{equation}
W(\alpha)\equiv\dfrac{1}{\pi^{2}}\int\int_{-\infty}^{+\infty}\exp(\beta^{*}\alpha-\beta\alpha^{*})C_{W}(\beta)d^{2}\beta.
\end{equation}
where $C_{W}(\lambda)$ is the symmetrically ordered characteristic
function, defined as: 

\begin{equation}
C_{W}(\lambda)=Tr\left[\rho e^{\lambda\hat{a}^{\dagger}-\lambda^{*}\hat{a}}\right].
\end{equation}
Here, we use the notation $\lambda^{\prime}$ and $\lambda^{\prime\prime}$
for the real and imaginary parts of $\lambda$, and set $\alpha=x+ip$
to emphasize the analogy between the quadratic radiation field and
the normalized dimensionless position and momentum observables of
the beam in phase space. We can rewrite the definition of the Wigner
function in terms of $x,p$ and $\lambda^{\prime},\lambda^{\prime\prime}$
as: 

\begin{equation}
W(x,p)=\frac{1}{\pi^{2}}\int\int_{-\infty}^{+\infty}e^{2i(p\lambda^{\prime}-x\lambda^{\prime\prime})}C_{W}(\lambda)d\lambda^{\prime}d\lambda^{\prime\prime}.\label{eq:wigner-function}
\end{equation}

By substituting the final normalized pointer state $\vert\Psi\rangle$
into Eq. (\ref{eq:wigner-function}), we obtain the explicit expression
for $W(x,p)$. Due to its complexity, we have provided its full expression
in the Appendix \ref{sec:wigner-app}. To illustrate the effects of
the postselected von Neumann measurement on the non-classical features
of $\vert\Psi\rangle$ in phase space (SPACS), we plot the corresponding
Wigner function in Fig. \ref{fig:wigner-function} for different values
of $r$ and the coupling strength $\Gamma$, with a fixed weak value
of $\langle\sigma_{x}\rangle_{w}=5.671(\alpha=\frac{8\pi}{9})$.

Each column, from left to right, corresponds to different coupling
strength parameters $\Gamma=0,\,0.3,\,1$, while each row, from top
to bottom, represents different values of $r=2$. The positive peak
of the Wigner function shifts from the center to the edge of phase
space, and its shape gradually becomes more irregular as $\Gamma$
increases. From the first column (see Fig. \ref{fig:wigner-function}),
we can observe that the original state $\vert\Psi^{\prime}\rangle$
is a Gaussian state, and its Wigner function remains positive. In
the second to fourth columns, corresponding to nonzero values of $\Gamma$,
we see the phase space density function $W\left(z\right)$ after the
postselected von Neumann measurement. The distribution exhibits squeezing
in phase space compared to the original state, and this squeezing
becomes more pronounced as $\Gamma$ increases.

Furthermore, Fig. \ref{fig:wigner-function} shows an increase in
the negative regions of the Wigner function as $\Gamma$ grows. The
presence of larger negative regions indicates a greater degree of
nonclassicality in the state. From this analysis, we conclude that
after the postselected von Neumann measurement, the phase space distribution
of $\vert\Psi\rangle$ not only becomes squeezed but also exhibits
enhanced nonclassicality for appropriate parameter values. 
\begin{flushleft}
\begin{figure*}
\begin{centering}
\includegraphics[width=18cm]{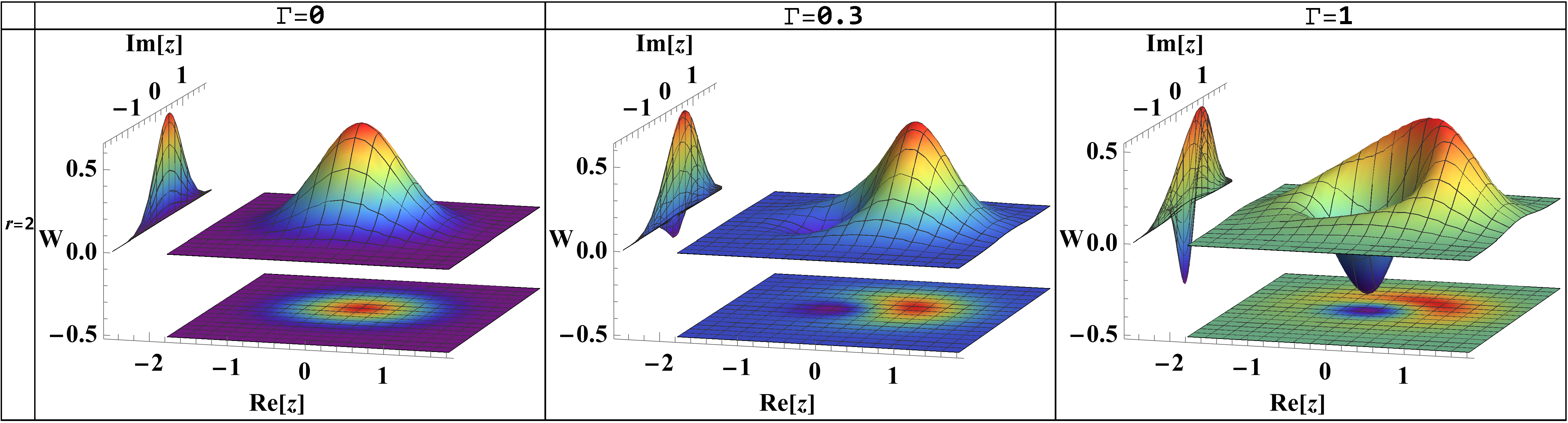}
\par\end{centering}
\caption{Wigner function of $\vert\Psi\rangle$ for different system parameters.
Here, we set $\varphi=0$, $\delta=0$, and $\alpha=\frac{8\pi}{9}$.\label{fig:wigner-function}}
\end{figure*}
\par\end{flushleft}

\section{Signal-To-Noise Ratio And Fidelity\label{VI}}

\subsection{Signal-To-Noise Ratio}

In precision measurement, obtaining precise information while suppressing
the associated noise is crucial. Previous studies \citep{Turek_2015}
have investigated the advantages of postselected von Neumann measurement
using HG and LG modes as pointers. These studies claimed that higher-order
modes of LG and HG beams do not offer any advantage in precision measurement
compared to the zero-mean fundamental Gaussian beam, even with large
anomalous weak values. The superposition of states can lead to interference
phenomena and produce results that single states cannot. To evaluate
the superiority of the state $\vert\Psi\rangle$ over the fundamental
Gaussian beam in postselected precision measurement, we explore the
SNR ratios between postselected and non-postselected measurements
\citep{Agarwal2013}: 

\begin{equation}
\chi=\frac{\mathcal{R}_{p}}{\mathcal{R}_{n}}.
\end{equation}
Here, $\mathcal{R}_{p}$ represents the SNR of the postselected von
Neumann measurement, which is defined by:

\begin{equation}
\mathcal{R}_{p}=\frac{\sqrt{NP_{s}}\left|\delta x\right|}{\triangle x}
\end{equation}
with the variance of position operator

\begin{equation}
\triangle x=\sqrt{\langle\Psi|\hat{X}^{2}|\Psi\rangle-\langle\Psi|\hat{X}|\Psi\rangle^{2}},
\end{equation}
and the average shift of the pointer variable $x$ after postselected
measurement

\begin{equation}
\delta x=\langle\Psi\vert\hat{X}|\Psi\rangle-\langle\Psi_{i}|\hat{X}|\Psi_{i}\rangle.
\end{equation}
Here, $\hat{X}=\sigma\left(\hat{a}+\hat{a}^{\dagger}\right)$is the
position operator, $N$ is the total number of measurements, and $P_{s}$
is the probability of finding the postselected state for a given preselected
state. For our scheme, this is given by $P_{s}=\vert\langle\psi_{f}\vert\psi_{i}\rangle\vert^{2}=\cos^{2}\frac{\alpha}{2}$,
where $NP_{s}$ represents the number of times the system is found
in the postselected state $\vert\psi_{f}\rangle$. By using the expressions
for the states $\vert\Psi_{i}\rangle$ and $\vert\Psi\rangle$ given
in Eq. (\ref{eq:4-1}) and Eq. (\ref{eq:4}) , we can obtain:

\begin{eqnarray}
\langle\Psi_{i}|\hat{X}|\Psi_{i}\rangle & = & 2\sigma Re\left[\langle\hat{a}\rangle\right],\\
\langle\Psi\vert\hat{X}\vert\Psi\rangle & = & 2\sigma Re\left[\langle\hat{a}\rangle\right],\\
\langle\Psi\vert\hat{X}^{2}\vert\Psi\rangle & = & \frac{\sigma^{2}}{2}\left\{ \langle\hat{a}^{\dagger}\hat{a}\rangle+Re\left[\langle\hat{a}^{2}\rangle\right]+2\right\} ,
\end{eqnarray}
Furthermore, when dealing with non-postselected measurements, there
is no postselection process after the interaction between the system
and the pointer. Thus, the definition of SNR $\mathcal{R}_{n}$ for
non-postselected weak measurement can be given as: 

\begin{equation}
\mathcal{R}_{n}=\frac{\sqrt{N}\left|\delta x^{\prime}\right|}{\triangle x^{\prime}}.
\end{equation}
 with 

\begin{align}
\triangle x' & =\sqrt{\langle\text{\ensuremath{\Phi}}|\hat{X}^{2}|\text{\ensuremath{\Phi}}\rangle-\langle\text{\ensuremath{\Phi}}|\hat{X}|\text{\ensuremath{\Phi}}\rangle^{2}},\\
\delta x' & =\langle\text{\ensuremath{\Phi}}|\hat{X}|\text{\ensuremath{\Phi}}\rangle-\langle\Psi_{i}|\hat{X}|\Psi_{i}\rangle.
\end{align}
Here, $\langle\text{\ensuremath{\Phi}}|\hat{X}|\text{\ensuremath{\Phi}}\rangle$
denotes the expectation value of the measuring observable under the
final state of the pointer without postselection, which can be derived
from Eq. (\ref{eq:3-1}). To evaluate the ratio $\chi$ of SNRs, we
need to calculate the related quantities, and the corresponding expressions
are given as:

\begin{eqnarray}
\langle\Psi_{i}\vert\hat{X_{1}}\vert\Psi_{i}\rangle & = & 2\sigma Re\left[\langle\hat{a}\rangle\right],\\
\langle\text{\ensuremath{\Phi}}\vert\hat{X_{1}}\vert\text{\ensuremath{\Phi}}\rangle & = & 2\sigma Re\left[\langle\hat{a}\rangle\right],\\
\langle\text{\ensuremath{\Phi}}\vert\hat{X}_{1}^{2}\vert\text{\ensuremath{\Phi}}\rangle & = & \frac{\sigma^{2}}{2}\left\{ \langle\hat{a}^{\dagger}\hat{a}\rangle+Re\left[\langle\hat{a}^{2}\rangle\right]+2\right\} ,
\end{eqnarray}
where the expectation values of $\langle\hat{a}\rangle$, $\langle\hat{a}^{\dagger}\hat{a}\rangle$
and $\langle\hat{a}^{2}\rangle$ under the state $\vert\Phi\rangle$
are given as

\begin{equation}
\langle\hat{a}\rangle=\frac{\gamma\mathrm{e}^{\mathrm{i}\varphi}}{\sqrt{2}(1+\gamma^{2})}+\frac{\Gamma}{2}\sin\alpha\cos\delta,
\end{equation}

\begin{eqnarray}
\langle\hat{a}^{\dagger}\hat{a}\rangle & = & \frac{\Gamma^{2}}{4}+\frac{\gamma^{2}}{2(1+\gamma^{2})},\\
\langle\hat{a}^{2}\rangle & = & \frac{\Gamma^{2}}{4}+\frac{\Gamma\gamma\mathrm{e}^{\mathrm{i}\varphi}\left(1+\sin\alpha\cos\delta\right)}{\sqrt{2}(1+\gamma^{2})},
\end{eqnarray}

\begin{figure}
\begin{centering}
\includegraphics[width=8cm]{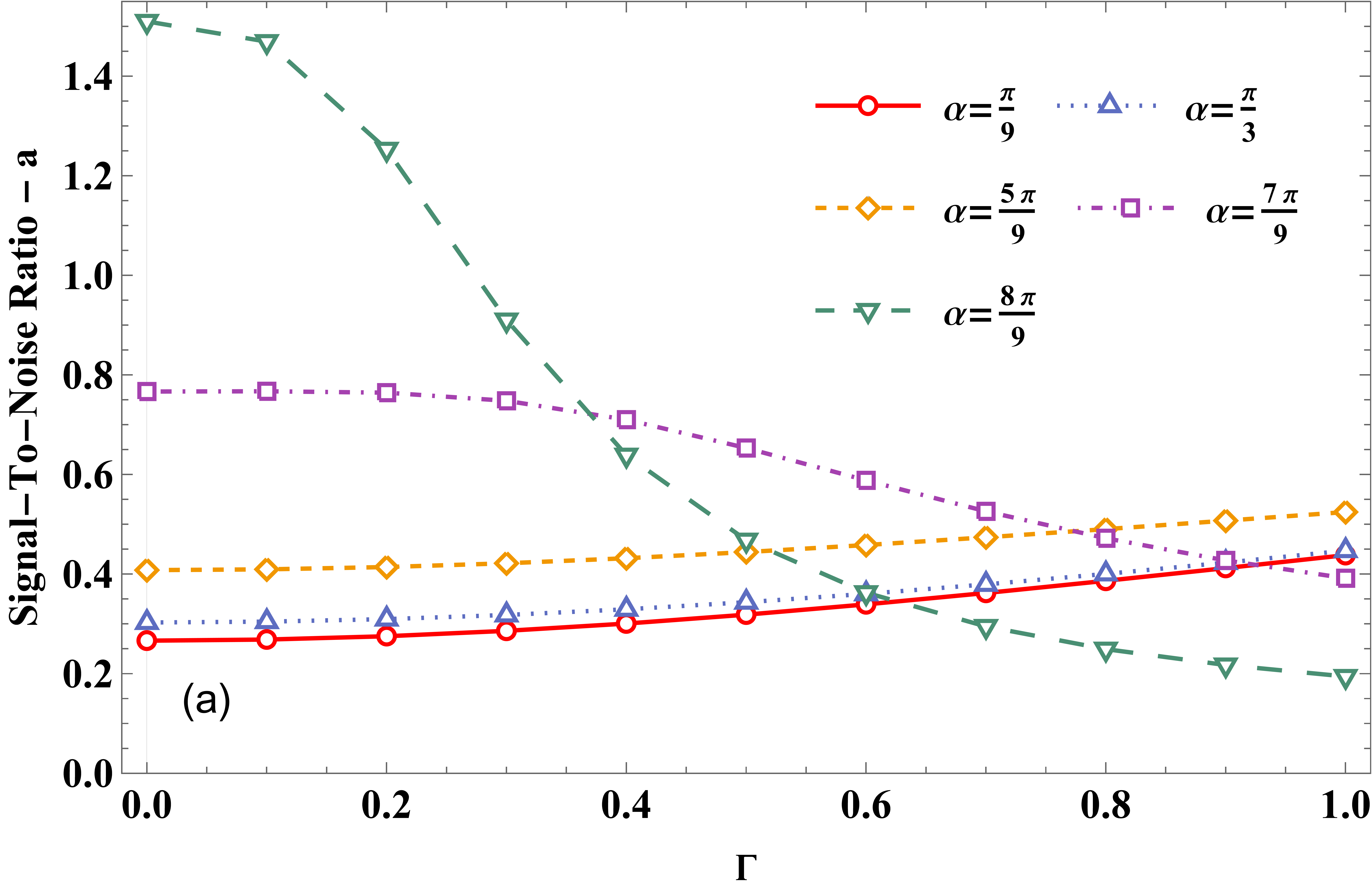}
\par\end{centering}
\begin{centering}
\includegraphics[width=8cm]{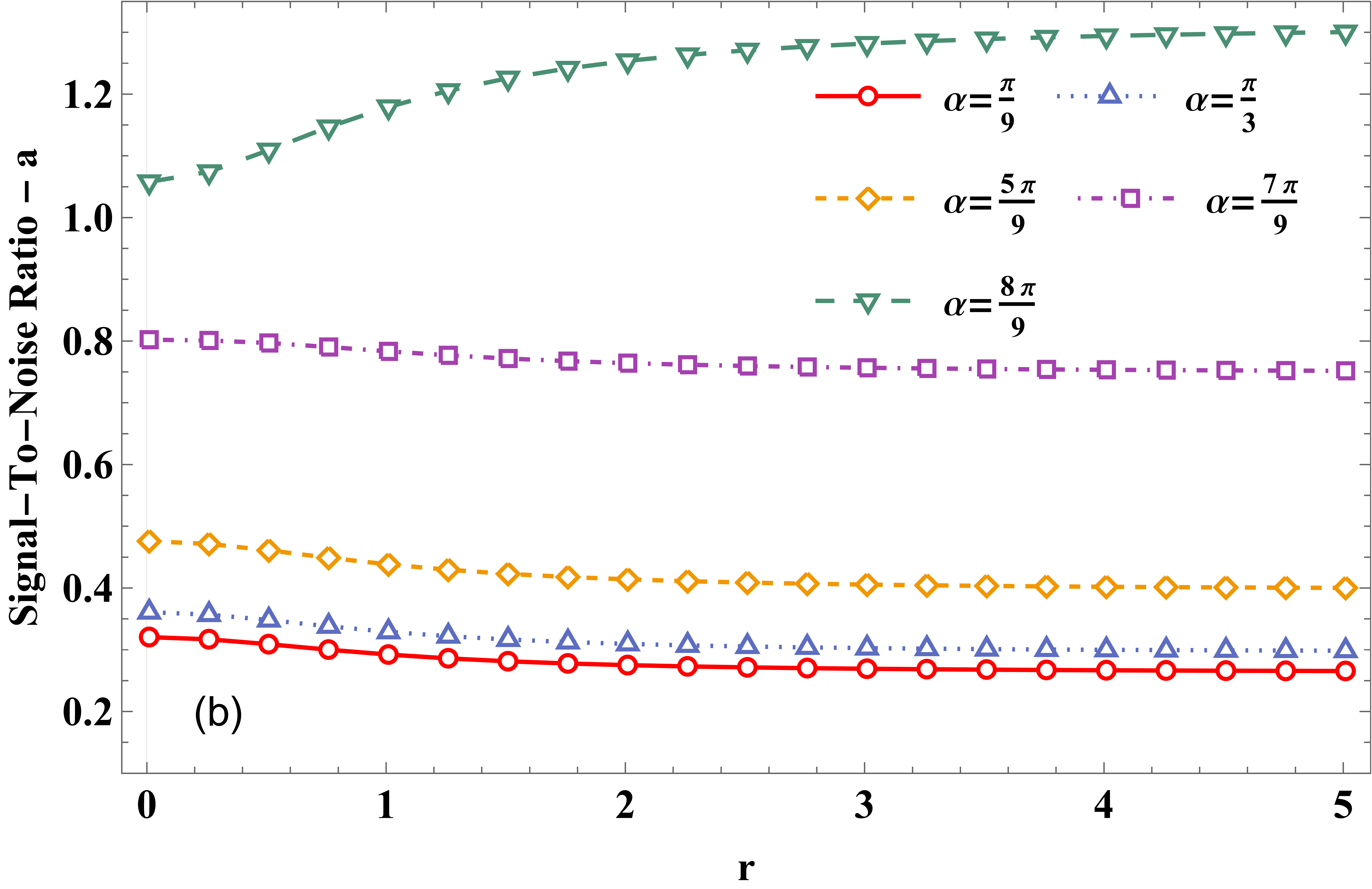}
\par\end{centering}
\begin{centering}
\includegraphics[width=8cm]{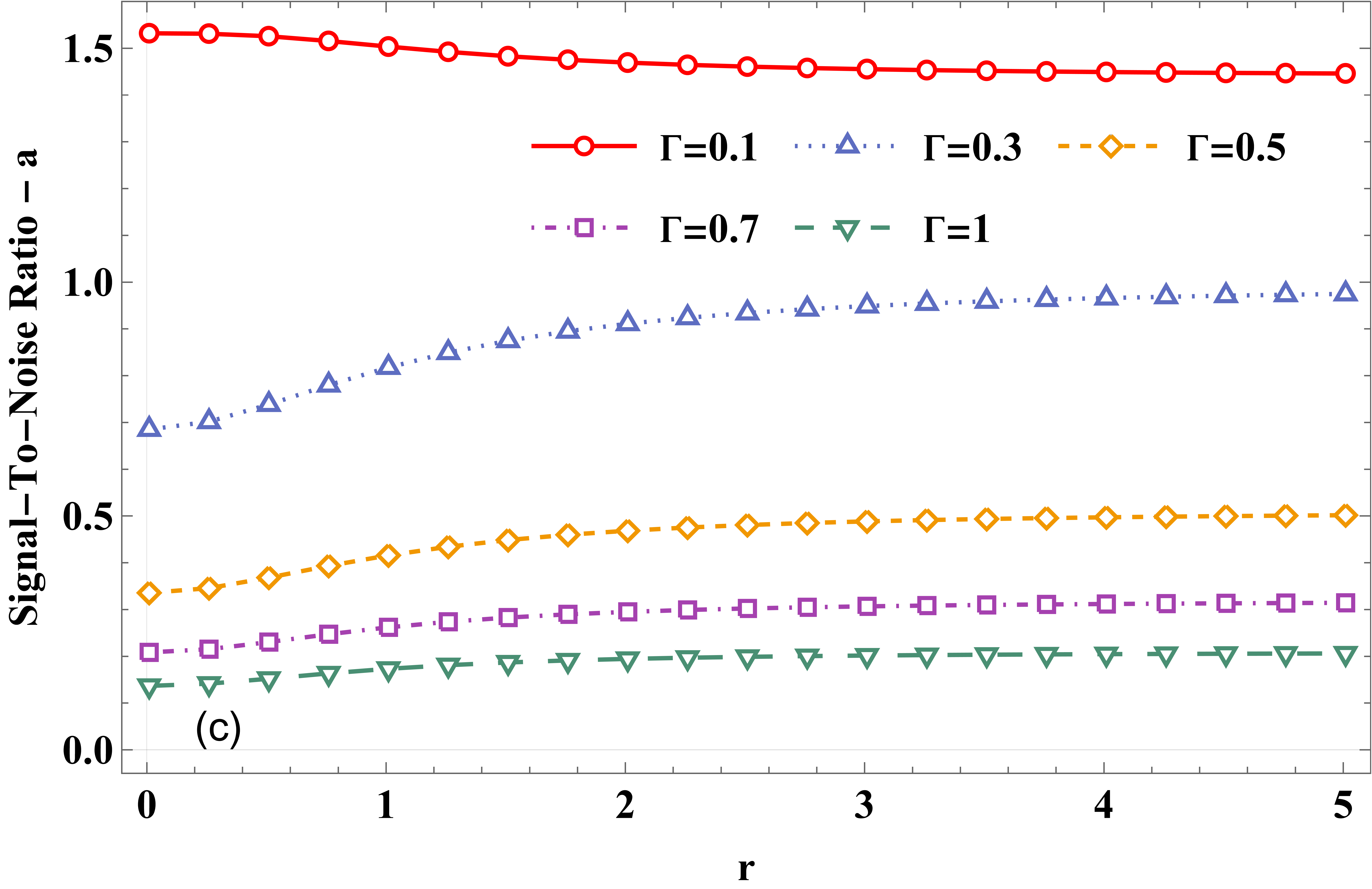}
\par\end{centering}
\caption{(a) The ratio $\chi$ as a function of the coupling strength $\Gamma$
for different weak values, with $r=2$. (b) The ratio $\chi$ as a
function of r for different weak values, with the coupling strength
at $\Gamma=0.2$ . (c) The ratio $\chi$ as a function of r for different
coupling strength $\Gamma$ , with $r=2$ . Here, we set $\delta=0$
and $\varphi=\frac{\pi}{2}$ . \label{fig:SNR}}
\end{figure}

The ratio of SNRs between postselected and non-postselected weak measurement
is plotted as a function of the coupling strength parameter $\Gamma$
and $r$, respectively, with the results shown in Figs.\textcolor{blue}{\ref{fig:SNR}}(a)
and \textcolor{blue}{\ref{fig:SNR}}(b). As observed in Fig.\textcolor{blue}{\ref{fig:SNR}}(a),
the ratio $\chi$ increases and can become greater than unity with
increasing weak values in the weak measurement regime. In Fig.\textcolor{blue}{\ref{fig:SNR}}(b),
we plot the ratio $\chi$ as a function of $r$ for different weak
values while fixing the coupling parameter at $\Gamma=0.2$. The numerical
results indicate that, in our scheme, anomalous weak values are indeed
helpful for increasing the SNR in weak measurement regimes. 

In Fig.\ref{fig:SNR}(c), we plot the ratio $\chi$ as a functions
of $r$ for different coupling parameter while fixing the weak value
fixed at $\alpha=\frac{8\pi}{9}$ , We can clearly observe that within
a larger range of weak values, there is a significant difference between
signals that have low coupling but high signal-to-noise ratios, and
those with high coupling but relatively lower signal-to-noise ratios.
Especially in the environment of weak coupling, the signal-to-noise
ratio is significantly higher than in strong coupling areas. Based
on this observation, we can logically infer that under weak coupling
conditions, the signal-to-noise ratio demonstrates much more efficient
performance.

From this result, we can deduce that postselected measurements improve
the precision of measurement compared to the non-postselected case.
Additionally, our superposition state $\vert\Psi_{i}\rangle$ can
be used in the postselected von Neumann process considering large
weak values of the measured observable.

There is a close relationship between state fidelity and SNR. High
state fidelity generally leads to high SNR, and vice versa. Therefore,
in quantum measurement and quantum state engineering, improving state
fidelity is one of the important means to enhance SNR and measurement
accuracy,Then, next, we will use SNR to conduct a detailed explanation
and illustration.

\subsection{Fidelity}

\begin{figure}
\begin{centering}
\includegraphics[width=8cm]{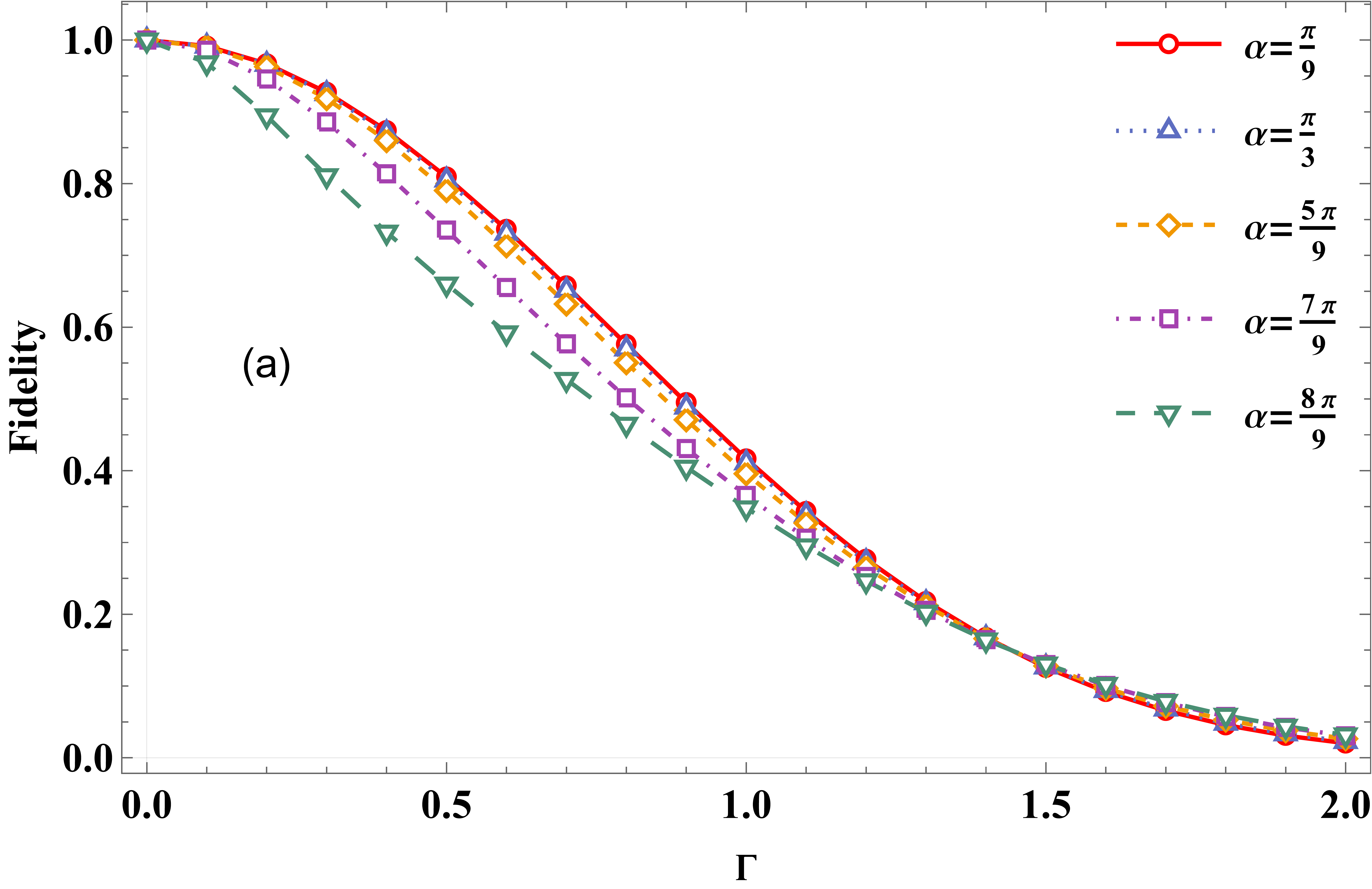}
\par\end{centering}
\begin{centering}
\includegraphics[width=8cm]{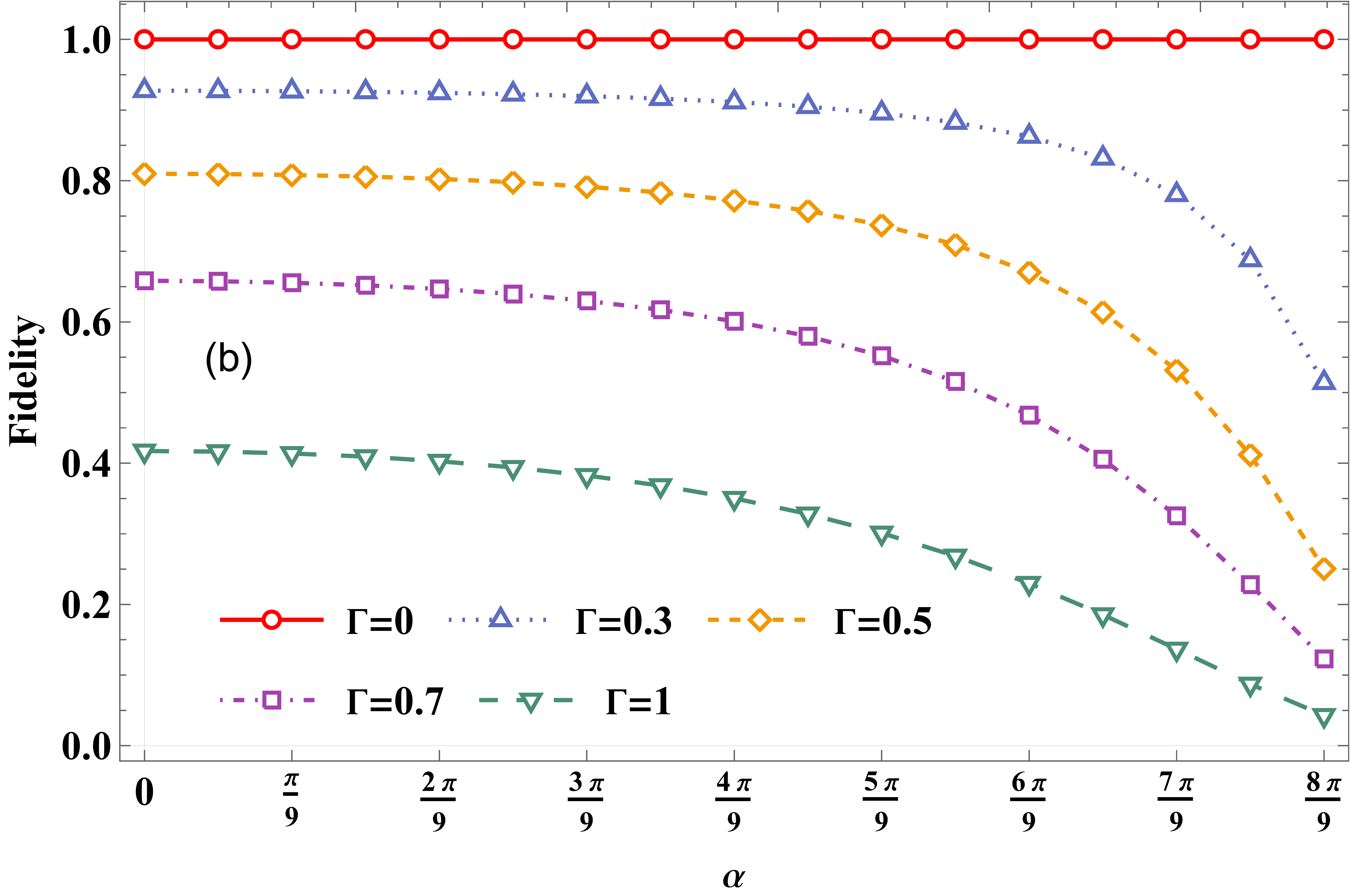}
\par\end{centering}
\caption{(a) Fidelity as a function of $\Gamma$ for different weak values
. (b) Fidelity as a function of weak values characterized by $\alpha$
for different $\Gamma$ . Here, we set $\delta=0$ , $r=2$ and $\varphi=\frac{\pi}{2}$
. \label{fig:Fidelity}}
\end{figure}

As discussed in above context, the inherent properties of the Superpositions
of the Gaussian and the Laguerre--Gaussian states are changed dramatically
after taking postselected von Neumann measurement. As shown in previous
studies, the postselected measurement could change the state, and
it may also be a reason to change the nonclassical features of original
given state. In this section, to examine the effects of postselected
measurement on the initial state $\vert\Psi\rangle$, we check the
state distance. In quantum information theory the distance between
two quantum states described by density operators $\rho$ and $\sigma$,
is defined as $F=\left(Tr\sqrt{\sqrt{\rho}\sigma\sqrt{\rho}}\right)^{2}$
, This formula also can characterize the fidelity (and is also called
Uhlmann-Jozsa fidelity) between two states. In our present study both
states are pure i.e., $\rho=\vert\Psi_{i}\rangle\langle\Psi_{i}\vert$
and $\sigma=\vert\Psi\rangle\langle\Psi\vert$, then the above formula
can be rewritten as

\begin{align}
F & =\left|\langle\Psi_{i}|\Psi\rangle\right|^{2}\nonumber \\
 & =\left|\frac{\lambda}{2}\left[\left(1-\langle\sigma_{x}\rangle_{w}\right)I_{2}+\left(1+\langle\sigma_{x}\rangle_{w}\right)I_{1}\right]\right|^{2}.
\end{align}
Its value is bounded $0\leq F\leq1$. If $F=1(F=0)$, then the two
states are exactly the same (totally different). This quantity is
a natural candidate for the state distance because it corresponds
to the closeness of states in the Hilbert space. Where

\begin{equation}
I_{2}=\frac{1}{1+\gamma^{2}}\left[1+\gamma^{2}+i\sqrt{2}\Gamma\gamma\sin(\varphi)-\frac{\gamma^{2}\Gamma^{2}}{2}\right]e^{-\frac{\Gamma^{2}}{2}}.
\end{equation}
Fidelity of the Superpositions of the Gaussian and the Laguerre--Gaussian
states under $\vert\Psi\rangle$ for different system parameters.
As illustrated in Figs.\ref{fig:Fidelity}(a) and (b), the post-selected
von Neumann measurement indeed induces a change in the given state.
As the coupling strength parameter($\Gamma$) and the parameter($r$)
increase, the distance between the states $|\Psi_{i}\rangle$ and
$|\Psi\rangle$ gradually expands, eventually leading to a distinguishable
state, regardless of the magnitude of the weak values. However, larger
weak values have a more pronounced effect on the distinct transformation
between states.

\section{Conclusion and outlooks\label{VII}}

As investigated in Ref. \citep{2020Q}, postselected measurements
have demonstrated advantages in quantum metrology and are particularly
useful in precision measurement processes. In this work, our results
based on the superposition of LG pointer states further confirm the
effectiveness of postselected measurements in state optimization and
precision measurements. We explored the effects of postselected von
Neumann measurements on the properties of superpositions of Gaussian
and Laguerre-Gaussian states. Throughout the work, we avoided approximations,
providing precise analytical results for all related quantities. First,
we presented analytical and numerical results for the quadrature squeezing
of the final pointer state. Our findings indicate that quadrature
squeezing increased significantly in the weak measurement regime for
anomalous weak values. By analyzing the second-order cross-correlation
function, we found no quantum correlation between the two modes of
the state $\vert\Psi_{i}\rangle$, even when signal amplification
via large anomalous weak values was employed. We also investigated
the phase space distribution of the final pointer state $\vert\Psi\rangle$.
While the initial state $\vert\Psi_{i}\rangle$ is a typical Gaussian
state, its Gaussianity changed after applying postselected measurements.
Specifically, the negativity of the Wigner function for the superpositions
of Gaussian and Laguerre-Gaussian states increased with the coupling
strength parameter and weak values. In the phase space distribution,
squeezing in two quadratures was also observed after the postselected
measurement. These results demonstrate that postselected von Neumann
measurements can transform the state $\vert\Psi_{i}\rangle$ from
a classical state to a nonclassical one by selecting appropriate coupling
strength parameters and weak values of the measured system's observables.
Furthermore, comparing the signal-to-noise ratio between postselected
and non-postselected measurement schemes shows that the SNR in the
postselected measurement scheme is significantly higher than in the
non-postselected scheme within the weak measurement regime, particularly
for large weak values. Despite the low probability of successful postselection,
the amplification effect of weak values plays a crucial role throughout
this work. This study contributes to the domain of state optimization
using postselected measurements. However, the most widely used quantum
state engineering method in quantum optics remains the addition or
subtraction of single photons to/from a light field \citep{2007PV}.
This approach to state optimization and manipulation can be applied
to single and multimode quantum states of light \citep{2016AV,PhysRevA.110.033717}.
Beyond photon addition and subtraction, or their superpositions, quantum
catalysis is also a feasible method for generating nonclassical quantum
states \citep{RN5}. Compared to these existing state optimization
techniques, the postselected von Neumann measurement-based method
offers a universal approach applicable to diverse quantum states.
A von Neumann-type interaction Hamiltonian can always be constructed
by selecting two independent degrees of freedom from the associated
states. Consequently, it would be interesting to investigate the effects
of postselected von Neumann measurements on other Gaussian and non-Gaussian
multipartite continuous-variable radiation fields \citep{2020quantum,PRXQuantum.2.030204,PhysRevA.109.040101},
including two-photon coherent states \citep{Agarwal:88,PhysRevA.50.2865,PhysRevA.13.2226},
three-mode states \citep{1990,PhysRevA.64.052303}and other multimode
radiation fields \citep{PhysRevA.42.4102}.
\begin{acknowledgments}
This work is supported by the National Natural Science Foundation
of China (Grants No. 12365005 and No. 11865017).
\end{acknowledgments}

\appendix

\section{\label{sec:A1}Relevant expressions}

The expectation values of some relevant operators under the state
$\vert\Psi\rangle$ are listed below:\begin{widetext} 
\begin{align}
\langle\hat{a}\rangle & =\frac{\vert\lambda^{\prime}\vert^{2}}{2}\left\{ \left(1+\vert\langle\sigma_{x}\rangle_{w}\vert^{2}\right)\frac{\gamma\mathrm{e}^{\mathrm{i}\varphi}}{\sqrt{2}(1+\gamma^{2})}+\left(1-\vert\langle\sigma_{x}\rangle_{w}\vert^{2}\right)II+\Gamma\left(1-I_{2}\right)Re\left[\langle\sigma_{x}\rangle_{w}\right]\right\} ,
\end{align}

\begin{align}
\langle\hat{b}\rangle & =\frac{\vert\lambda^{\prime}\vert^{2}i\sqrt{2}\gamma\mathrm{e}^{\mathrm{i}\varphi}}{4(1+\gamma^{2})}\left[1+\vert\langle\sigma_{x}\rangle_{w}\vert^{2}+\left(1-\vert\langle\sigma_{x}\rangle_{w}\vert^{2}\right)e^{-\frac{\Gamma^{2}}{2}}\right]-\frac{i\vert\lambda^{\prime}\vert^{2}\gamma^{2}\Gamma}{2(1+\gamma^{2})}Im\left[\langle\sigma_{x}\rangle_{w}\right]e^{-\frac{\Gamma^{2}}{2}},
\end{align}

\begin{align}
\langle\hat{a}^{2}\rangle & =\frac{\vert\lambda^{\prime}\vert^{2}\Gamma}{2}\left[\left(\frac{\sqrt{2}\gamma\mathrm{e}^{\mathrm{i}\varphi}}{1+\gamma^{2}}+2II\right)Re\left[\langle\sigma_{x}\rangle_{w}\right]+\left(1+\vert\langle\sigma_{x}\rangle_{w}\vert^{2}\right)\frac{\Gamma}{4}\right]\nonumber \\
 & +\frac{\vert\lambda^{\prime}\vert^{2}\Gamma^{2}}{16}\left[\left(1+\langle\sigma_{x}\rangle_{w}^{\ast}\right)\left(1-\langle\sigma_{x}\rangle_{w}\right)I_{2}+\left(1-\langle\sigma_{x}\rangle_{w}^{\ast}\right)\left(1+\langle\sigma_{x}\rangle_{w}\right)I_{1}\right],
\end{align}

\begin{align}
\langle\hat{b}^{2}\rangle & =0,
\end{align}

\begin{align}
\langle\hat{a}^{\dagger}\hat{a}\rangle & =\frac{\vert\lambda^{\prime}\vert^{2}}{2}\left(1+\vert\langle\sigma_{x}\rangle_{w}\vert^{2}\right)\left(\frac{\gamma^{2}}{2(1+\gamma^{2})}+\frac{\Gamma^{2}}{4}\right)+i\frac{\vert\lambda^{\prime}\vert^{2}\Gamma\gamma\cos(\varphi)}{2\sqrt{2}(1+\gamma^{2})}Im\left[\langle\sigma_{x}\rangle_{w}\right]+\frac{\vert\lambda^{\prime}\vert^{2}}{4}\left(1+\langle\sigma_{x}\rangle_{w}^{\ast}\right)\left(1-\langle\sigma_{x}\rangle_{w}\right)III_{+}\nonumber \\
 & +\frac{\vert\lambda^{\prime}\vert^{2}}{4}\left(1-\left|\langle\sigma_{x}\rangle_{w}^{\ast}\right|^{2}\right)III_{-}+\frac{\vert\lambda^{\prime}\vert^{2}\Gamma^{2}}{16}\left[\left(1+\langle\sigma_{x}\rangle_{w}^{\ast}\right)\left(1-\langle\sigma_{x}\rangle_{w}\right)I_{2}+\left(1-\langle\sigma_{x}\rangle_{w}^{\ast}\right)\left(1+\langle\sigma_{x}\rangle_{w}\right)I_{1}\right]\nonumber \\
 & +\frac{\vert\lambda^{\prime}\vert^{2}\Gamma}{8}\left[\left(1-\langle\sigma_{x}\rangle_{w}^{\ast}\right)\left(1+\langle\sigma_{x}\rangle_{w}\right)\left(IV_{1}+II\right)-\left(1+\langle\sigma_{x}\rangle_{w}^{\ast}\right)\left(1-\langle\sigma_{x}\rangle_{w}\right)\left(IV_{2}+II\right)\right],
\end{align}

\begin{align}
\langle\hat{b}^{\dagger}\hat{b}\rangle & =\frac{\vert\lambda^{\prime}\vert^{2}}{4}\left[\left(1+\vert\langle\sigma_{x}\rangle_{w}\vert^{2}\right)\frac{\gamma^{2}}{1+\gamma^{2}}+\left(1-\vert\langle\sigma_{x}\rangle_{w}\vert^{2}\right)\frac{\gamma^{2}}{1+\gamma^{2}}e^{-\frac{\Gamma^{2}}{2}}\right],
\end{align}

\begin{align}
\langle\hat{a}^{\dagger}\hat{b}\rangle & =\frac{\vert\lambda^{\prime}\vert^{2}}{4}\Bigg[\left(1+\vert\langle\sigma_{x}\rangle_{w}\vert^{2}\right)\frac{i\gamma^{2}}{(1+\gamma^{2})}+Im\left[\langle\sigma_{x}\rangle_{w}\right]\frac{i\Gamma\gamma\mathrm{e}^{\mathrm{i}\varphi}}{\sqrt{2}(1+\gamma^{2})}\left(1+e^{-\frac{\Gamma^{2}}{2}}\right)+\left(1-\vert\langle\sigma_{x}\rangle_{w}\vert^{2}\right)\nonumber \\
 & \times\frac{\gamma^{2}\Gamma^{2}e^{-\frac{\Gamma^{2}}{2}}}{2(1+\gamma^{2})}\Bigg]+\frac{\vert\lambda^{\prime}\vert^{2}}{4}\left[\left(1+\langle\sigma_{x}\rangle_{w}^{\ast}\right)\left(1-\langle\sigma_{x}\rangle_{w}\right)B_{+}+\left(1-\langle\sigma_{x}\rangle_{w}^{\ast}\right)\left(1+\langle\sigma_{x}\rangle_{w}\right)B_{-}\right],
\end{align}

\begin{align}
\langle\hat{a}\hat{b}\rangle & =\frac{\vert\lambda^{\prime}\vert^{2}\gamma\Gamma}{8\left(1+\gamma^{2}\right)}\left\{ 2\sqrt{2}i\mathrm{e}^{\mathrm{i}\varphi}\left(Re\left[\langle\sigma_{x}\rangle_{w}\right]+iIm\left[\langle\sigma_{x}\rangle_{w}\right]e^{-\frac{\Gamma^{2}}{2}}\right)+\left(1-\vert\langle\sigma_{x}\rangle_{w}\vert^{2}\right)\gamma\Gamma e^{-\frac{\Gamma^{2}}{2}}\right\} ,
\end{align}

\begin{align}
\langle\hat{a}^{\dagger}\hat{a}\hat{b}^{\dagger}\hat{b}\rangle & =\frac{\vert\lambda^{\prime}\vert^{2}\Gamma^{2}\gamma^{2}}{16(1+\gamma^{2})}\left[1+\vert\langle\sigma_{x}\rangle_{w}\vert^{2}-\left(1-\vert\langle\sigma_{x}\rangle_{w}\vert^{2}\right)e^{-\frac{\Gamma^{2}}{2}}\right],
\end{align}

\begin{align}
\langle\hat{a}^{\dagger2}\hat{a}^{2}\rangle & =\frac{\vert\lambda^{\prime}\vert^{2}}{4}\left[\left(1-\langle\sigma_{x}\rangle_{w}^{\ast}\right)\left(1-\langle\sigma_{x}\rangle_{w}\right)M_{-}+\left(1+\langle\sigma_{x}\rangle_{w}^{\ast}\right)\left(1+\langle\sigma_{x}\rangle_{w}\right)M_{+}\right]\nonumber \\
 & +\frac{\vert\lambda^{\prime}\vert^{2}}{4}\left[\left(1-\langle\sigma_{x}\rangle_{w}^{\ast}\right)\left(1+\langle\sigma_{x}\rangle_{w}\right)M_{1}+\left(1+\langle\sigma_{x}\rangle_{w}^{\ast}\right)\left(1-\langle\sigma_{x}\rangle_{w}\right)M_{2}\right],
\end{align}

\begin{align}
\langle\hat{b}^{\dagger2}\hat{b}^{2}\rangle & =0.
\end{align}

with 

\begin{align}
II & =\frac{\gamma\mathrm{e}^{\mathrm{i}\varphi}e^{-\frac{\Gamma^{2}}{2}}}{\sqrt{2}(1+\gamma^{2})},
\end{align}

\begin{align}
III_{\pm} & =\frac{e^{-\frac{\Gamma^{2}}{2}}}{1+\gamma^{2}}\left[\frac{\gamma^{2}}{2}\left(1-\Gamma^{2}\right)\pm\frac{\gamma\mathrm{e}^{\mathrm{i}\varphi}}{\sqrt{2}}\Gamma\right],
\end{align}

\begin{align}
B_{\pm} & =\frac{\pm i\gamma e^{-\frac{\Gamma^{2}}{2}}}{1+\gamma^{2}}\left[\frac{\gamma}{2}(1-\Gamma^{2})-\frac{\mathrm{e}^{\pm\mathrm{i}\varphi}}{\sqrt{2}}\Gamma\right],
\end{align}

\begin{align}
M_{\pm} & =\frac{\Gamma^{2}\gamma^{2}}{2(1+\gamma^{2})}\pm\frac{\Gamma^{3}\gamma\cos(\varphi)}{2\sqrt{2}(1+\gamma^{2})}+\frac{\Gamma^{4}}{16},
\end{align}

\begin{align}
M_{1} & =\Gamma T_{+}+\frac{\Gamma^{2}}{4}\left(T+4IV_{1}\right)+\frac{\Gamma^{3}}{4}\left(V_{+}+II\right)+\frac{\Gamma^{4}I_{1}}{16},
\end{align}

\begin{align}
M_{2} & =-\Gamma T_{-}\frac{\Gamma^{2}}{4}\left(T+4IV_{2}\right)-\frac{\Gamma^{3}}{4}\left(V_{-}+II\right)+\frac{\Gamma^{4}I_{2}}{16},
\end{align}

\begin{align}
T_{\pm} & =\frac{\gamma\mathrm{e}^{\mathrm{i}\varphi}\Gamma}{1+\gamma^{2}}\left[\frac{\Gamma}{\sqrt{2}}\mp\gamma\mathrm{e}^{\mathrm{-i}\varphi}\left(1-\frac{\Gamma^{2}}{2}\right)\right]e^{-\frac{\Gamma^{2}}{2}},
\end{align}

\begin{align}
T & =\frac{\Gamma^{2}}{1+\gamma^{2}}\left[1+\gamma^{2}\left(2-\frac{\Gamma^{2}}{2}\right)\right]e^{-\frac{\Gamma^{2}}{2}},
\end{align}

\begin{align}
IV_{1} & =\frac{e^{-\frac{\Gamma^{2}}{2}}}{1+\gamma^{2}}\left[\frac{\gamma^{2}}{2}\left(1-\Gamma^{2}\right)-\frac{\gamma\mathrm{e}^{\mathrm{i}\varphi}}{\sqrt{2}}\Gamma\right],
\end{align}

\begin{align}
IV_{2} & =\frac{e^{-\frac{\Gamma^{2}}{2}}}{1+\gamma^{2}}\left[\frac{\gamma^{2}}{2}\left(1-\Gamma^{2}\right)+\frac{\gamma\mathrm{e}^{\mathrm{i}\varphi}}{\sqrt{2}}\Gamma\right],
\end{align}

\begin{align}
V_{\pm} & =\frac{e^{-\frac{\Gamma^{2}}{2}}}{2(1+\gamma^{2})}\left[\Gamma^{2}\gamma\mathrm{e}^{\mathrm{i}\varphi}+\sqrt{2}\gamma\mathrm{e}^{\mathrm{-i}\varphi}\left(1-\Gamma^{2}\right)\mp2\Gamma\mp\Gamma\gamma^{2}\left(1+\frac{2-\Gamma^{2}}{\sqrt{2}}\right)\right],
\end{align}

\section{\label{sec:wigner-app}Wigner function}

The explicit expression of the Wigner function of our final pointer
state $\vert\Psi\rangle$ can be calculated as following: 
\begin{eqnarray*}
W(\alpha) & = & \frac{\vert\lambda^{\prime}\vert^{2}}{4}\Bigg\{\left|1-\langle\sigma_{x}\rangle_{w}\right|^{2}W_{+}+\left|1+\langle\sigma_{x}\rangle_{w}\right|^{2}W_{-}+2\Re\Big[\left(1+\langle\sigma_{x}\rangle_{w}^{\ast}\right)\left(1-\langle\sigma_{x}\rangle_{w}\right)W_{1}\Big]\Bigg\},
\end{eqnarray*}
where 

\begin{align}
W_{\pm} & =\frac{1}{\pi}\left\{ 2+\frac{2\gamma\sqrt{2}}{1+\gamma^{2}}\left[\left(2x\pm\Gamma\right)\cos(\varphi)+2p\sin(\varphi)\right]+\frac{\gamma^{2}}{1+\gamma^{2}}\left[4p^{2}+\left(2x\pm\Gamma\right)^{2}-2\right]\right\} e^{-2p^{2}-\frac{(2x\pm\Gamma)^{2}}{2}},\\
W_{1} & =\frac{e^{-\frac{\Gamma^{2}}{2}}}{\pi}\left\{ 2+\frac{4\gamma\sqrt{2}}{1+\gamma^{2}}\left[x\cos(\varphi)+p\sin(\varphi)\right]+\frac{2\gamma^{2}}{(1+\gamma^{2})}\left(2x^{2}+2p^{2}-1\right)\right\} e^{-2x^{2}-\frac{\left(2p-i\Gamma\right)^{2}}{2}},
\end{align}

\end{widetext}

\bibliographystyle{apsrev4-1}
\bibliography{Refs}

\end{document}